\documentclass[preprint, 3p]{elsarticle}

\usepackage{lineno,hyperref}
\modulolinenumbers[5]

\journal{Journal of Computational Physics}

\usepackage{amsmath}
\usepackage{float}
\usepackage{pgfplots}
\usepackage{subfig}
\usepackage{tabularx}
\newcolumntype{Y}{>{\centering\arraybackslash}X}
\usepackage{multirow}
\usepackage{verbatim}
\usepackage{amsfonts}
\usepackage{enumerate}
\usepackage{hyperref}
\usepackage{cancel}
\usepackage{soul}
\usepackage{xcolor}

\newenvironment{sbmatrix}[1]
 {\def\mysubscript{#1}\mathop\bgroup\begin{bmatrix}}
 {\end{bmatrix}\egroup_{\textstyle\mathstrut\mysubscript}}

\newcommand{\eqnref}[1]{Eq. \ref{#1}}
\newcommand{\figref}[1]{Fig. \ref{#1}}
\newcommand{\tableref}[1]{Table \ref{#1}}
\newcommand{\sectionref}[1]{Section \ref{#1}}

\begin{document}

\begin{frontmatter}

\title{A dimensionally split Cartesian cut cell method for hyperbolic conservation laws}

\author[lsc]{Nandan Gokhale\corref{cor1}}
\ead{nbg22@cam.ac.uk}
\author[lsc]{Nikos Nikiforakis}
\ead{nn10005@cam.ac.uk}
\author[fub]{Rupert Klein}
\ead{rupert.klein@math.fu-berlin.de}

\cortext[cor1]{Corresponding author}

\address[lsc]{Laboratory for Scientific Computing, Cavendish Laboratory,
University of Cambridge, Cambridge, CB3 0HE, UK}
\address[fub]{Institut f{\"u}r Mathematik, FB Mathematik und Informatik, Freie Universit{\"a}t Berlin, Arnimallee 6, 14195 Berlin, Germany}

\begin{abstract}
We present a dimensionally split method for solving hyperbolic conservation laws on Cartesian cut cell meshes. The approach combines local geometric and wave speed information to determine a novel stabilised cut cell flux, and we provide a full description of its three-dimensional implementation in the dimensionally split framework of Klein et al. \cite{Klein2009}. The convergence and stability of the method are proved for the one-dimensional linear advection equation, while its multi-dimensional numerical performance is investigated through the computation of solutions to a number of test problems for the linear advection and Euler equations. When compared to the cut cell flux of Klein et al., it was found that the new flux alleviates the problem of oscillatory boundary solutions produced by the former at higher Courant numbers, and also enables the computation of more accurate solutions near stagnation points. Being dimensionally split, the method is simple to implement and extends readily to multiple dimensions.
\end{abstract}


\end{frontmatter}

\section{Introduction}

\label{sect:Introduction}

Cartesian cut cell methods are a class of immersed boundary methods \cite{Mittal2005} which are designed to satisfy discrete conservation at the boundary. Specifically, the cut cell approach usually involves representing a boundary via a piecewise linear reconstruction in a Cartesian mesh, resulting in a sharp representation of the interface. The approach is an attractive alternative to conventional unstructured or body-fitted meshing techniques due to the ease of automatic mesh generation for complex geometries, and the computational conveniences offered by the use of Cartesian grids. However, the resulting mesh can contain cut cells of arbitrarily small volume fraction which impose severe constraints on the time step for explicit numerical schemes.

Different methodologies exist to overcome this `small cell problem'. With cell merging, as in Clarke et al. \cite{Clarke1986}, or the related cell linking strategy, as employed by Quirk \cite{Quirk1994}, Kirkpatrick et al. \cite{Kirkpatrick2003} and Hartmann et al. \cite{Hartmann2011}, cells with volume lower than a certain threshold are absorbed into larger neighbouring cells. Based on the static boundary cut cell formulation of Hartmann et al. \cite{Hartmann2011}, it may be noted that Schneiders et al. \cite{Schneiders2013} have successfully developed a method to compute moving boundary problems in 3D by introducing an interpolation routine and flux redistribution step. Meinke et al. \cite{Meinke2013} have used the developed technique to compute the flow in a relatively complicated engine geometry involving moving parts.

Another alternative is to use the simplified `$h$-box' method of Berger and Helzel \cite{Berger2012}. Here, the domain of dependence of the intercell flux is extended to the regular cell length, $h$, on a virtual grid of $h$-boxes in such a way that a `flux cancellation' occurs, removing the dependence on cell volume from the update formula. Although second order accuracy at the boundary can be achieved with this method, it is quite complicated and has not yet been implemented in three dimensions.

In the `flux redistribution' approach of Colella et al. \cite{Colella2006}, conservative but potentially unstable fluxes are initially computed for each cut cell. A stable but non-conservative part of the update is applied to the cut cells, and conservation is maintained by redistributing the remaining part to surrounding cut and uncut cells. A similar approach is the `flux mixing' technique of Hu et al. \cite{Hu2006}. Here, the explicit fluxes are used to update all cells, following which the solution in the cut cells is mixed with neighbouring cells. This technique has been extended and used in the context of a number of different applications. Grilli et al. \cite{grilli2012} have successfully used it at the compression ramp geometry in their study on shockwave turbulent boundary layer interaction. Pasquariello et al. \cite{pasquariello2016} use it at the interface in the coupled finite volume-finite element method they develop to study fluid-structure interaction problems. Furthermore, it may be noted that Muralidharan and Menon \cite{Muralidharan2016} have recently managed to use it to develop a third order accurate method.

Another noteworthy approach to obtain high order convergence is the `inverse Lax-Wendroff' method of Tan and Shu \cite{Tan2010}. In a subsequent paper describing an efficient implementation of the technique, Tan et al. \cite{Tan2012} are able to demonstrate fifth order convergence for two-dimensional problems for the Euler equations.

`Explicit-implicit' approaches as developed by Jebens et al. \cite{Jebens2012}, or more recently, by May and Berger \cite{May2014}, are also an interesting area of development. Here, the attempt is to develop a combined scheme where regular and cut cells are integrated explicitly and implicitly in time respectively.

All the aforementioned methods have their own relative merits and are implemented in an unsplit fashion. We are particularly interested in adopting a dimensionally split approach which is a simple way to extend one-dimensional methods for hyperbolic conservation laws to multidimensional problems. To that end, we make use of the framework introduced by Klein, Bates and Nikiforakis \cite{Klein2009} which provides a description of how cut cell updates can be performed in a split fashion. The stabilised `KBN' cut cell flux that they devise makes use of local geometric information. In this paper, we present a `Localised Proportional Flux Stabilisation' (LPFS) approach which makes use of local geometric and wave speed information to define a novel cut cell flux. Numerical tests indicate that the LPFS flux alleviates the problem of oscillatory boundary solutions produced by the KBN flux at higher Courant numbers, and enables the computation of more accurate solutions near stagnation points.

The rest of the paper is organised as follows. In \sectionref{sect:Governing_equations_and_solution_framework}, we outline the governing equations and explicit numerical schemes that we use. A summary of how the various cut cell geometric parameters are calculated is provided in \sectionref{sect:Cut_cell_mesh_generation}. In \sectionref{sect:Numerical_method}, we describe the derivation of the KBN and LPFS fluxes, and describe their multi-dimensional extensions in the framework of Klein et al. In \sectionref{sect:Convergence_and_stability_analysis}, a theoretical convergence and stability analysis of the LPFS method for the model one-dimensional linear advection equation is presented. In \sectionref{sect:Results}, we present numerical solutions for a number of multi-dimensional test problems to demonstrate the performance of the LPFS method. Finally, conclusions and areas for future work are provided in \sectionref{sect:Conclusions}.

\section{Governing equations and solution framework}

\label{sect:Governing_equations_and_solution_framework}

We use the linear advection equation,
\begin{equation}
\label{eqn:linAdv_eqns}
\partial_{t} u + \mathbf{a} \cdot \nabla u = 0,
\end{equation}
when proving the convergence and stability of LPFS in \sectionref{sect:Convergence_and_stability_analysis}, and for some convergence tests in \sectionref{sect:Convergence_tests}. $u$ is the variable being advected at constant velocity $\mathbf{a}$.

For more challenging tests, we solve the compressible, unsteady, Euler equations
\begin{align}
\label{eqn:eul_eqns}
\partial_{t} \rho + \nabla \cdot (\rho \mathbf{u}) &= 0,\nonumber\\
\partial_{t} (\rho \mathbf{u}) + \nabla \cdot (\rho \mathbf{u} \otimes \mathbf{u} + p I) &= \mathbf{0},\\
\partial_{t} E + \nabla \cdot [(E + p)\mathbf{u}] &= 0. \nonumber
\end{align}

In \eqnref{eqn:eul_eqns}, $\rho$ is density, $\mathbf{u}$ is velocity, $p$ is pressure and $I$ is the identity matrix. $E$ is the total energy per unit volume, given by
\begin{equation}
\label{eqn:TotE}
E = \rho \left(\frac{1}{2} |\mathbf{u}|^2 + e\right),
\end{equation}
where $e$ is the specific internal energy. To close the system of equations \eqnref{eqn:eul_eqns}-\eqnref{eqn:TotE} we use the ideal gas equation of state
\begin{equation}
\label{eqn:idealEOS}
e = \frac{p}{\rho (\gamma - 1)},
\end{equation}
where $\gamma$, the heat capacity ratio, is assumed to be 1.4.

We compute explicit fluxes in a Godunov-based finite volume framework using an exact Riemann solver and the MUSCL-Hancock scheme in conjunction with the van-Leer limiter \cite{Toro2009}. This scheme is second order accurate in smooth regions. The definition of the KBN and LPFS fluxes are independent of the particular choice of flux method used, however. We also make use of hierarchical AMR \cite{berger1984adaptive} to refine areas of interest such as shock waves, or the cut cell interface, while allowing the use of coarser resolutions elsewhere for the sake of computational efficiency. Multi-dimensional updates are performed using Strang splitting \cite{strang1968construction} in 2D, and straightforward Godunov splitting \cite{leveque2002finite} in 3D.

The time step, $\Delta t$, is restricted by the CFL condition
\begin{equation}
\label{eqn:cfl}
\Delta t = C_\text{cfl} \min_{d,i} \left( \frac{\Delta x_{d,i}}{W^\text{max}_{d,i}} \right),
\end{equation}
where $d$ is the index of the coordinate direction, $i$ is the index of a computational cell, and $\Delta x_{d,i}$ and $W^\text{max}_{d,i}$ are the spatial resolution and max wave speed for cell $i$ in the $d$ direction respectively. The MUSCL-Hancock scheme has a linearised stability constraint $C_\text{cfl} \in (0,1]$ \cite{Toro2009}, where $C_\text{cfl}$ is the Courant number.

For the Euler equations, the wave speed for cell $i$ in the $d$ direction, $W_{d,i}$, is computed using the following estimate suggested by Toro \cite{Toro2009}:
\begin{equation}
\label{eqn:wavespeeds}
W_{d,i} = |\mathbf{u}_{d,i}| + a_i,
\end{equation}
where $\mathbf{u}_{d,i}$ is the component of the velocity in cell $i$ in the $d$ direction. $a_i$ is the speed of sound in cell $i$, given by
\begin{equation}
\label{eqn:soundspeed}
a_i = \sqrt{\frac{\gamma p_i}{\rho_i}}.
\end{equation}

\section{Cut-cell mesh generation}

\label{sect:Cut_cell_mesh_generation}

In this section, we describe the calculation of the various cut cell geometric parameters.

\begin{figure}
\centering
\input{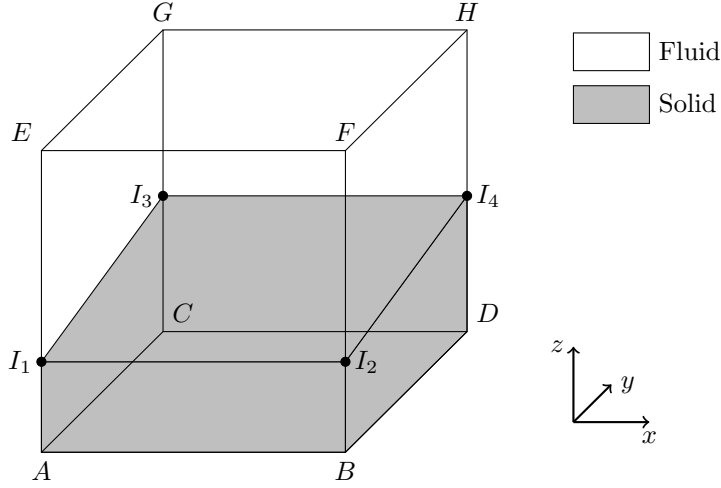}
\caption{Illustration of a cut cell in 3D.}
\label{fig:cutCell}
\end{figure}

Consider \figref{fig:cutCell}, which shows a 3D cut cell where the solid-fluid interface intersects the cell at four points $I_1$, $I_2$, $I_3$ and $I_4$. The following parameters are to be calculated:
\begin{enumerate}[(i)]
\item The coordinates of the intersection points.
\item The face fraction, $\beta \in [0,1]$, of each cell face. This represents the fluid area of the face non-dimensionalised by total cell face area.
\item The area, $A^b$, of the reconstructed interface in the cell and $\mathbf{\hat{n}}^b$, the interface unit normal. The superscript $b$ is short for `boundary'.
\item The volume fraction, $\alpha \in [0,1]$, of the cell. This is the fluid volume of the cell non-dimensionalised by the total cell volume.
\end{enumerate}

We proceed by treating the interface implicitly as the zero level-set of a signed distance function. The technique of Mauch \cite{Mauch2000} is used to compute the signed distance function, $\phi(\mathbf{x})$, at the vertices of the cell, and this information is used to reconstruct all the geometric information needed for evaluations of the flux balances.

Consider edge $AE$ in \figref{fig:cutCell}. If it is intersected by the interface, i.e., if the signed distances at points $A$ and $E$ are of opposite sign, then assuming a linear interface in the vicinity of the edge, the non-dimensional distance from $A$ to $I_1$ may be calculated to be
\begin{equation}
\label{eqn:SDIntersection}
\frac{AI_1}{AE} = -\left(\frac{\phi_A}{\phi_E - \phi_A}\right),
\end{equation}
where $\phi_A$ and $\phi_E$ are the signed distances at vertices $A$ and $E$ respectively. Clearly, this approach can be used to determine the coordinates of all the intersection points. Note that in a 1D simulation, this calculation directly gives the volume fraction of the cell. In a 2D simulation, it would be used to calculate the cell face fractions.

With the coordinates of the intersection points identified, calculating the face fractions for each face in 3D is a matter of making use of the formula for the area of an arbitrary non-self-intersecting polygon with $n$ ordered vertices $(x_1,y_1),...,(x_n,y_n)$ \cite{Weisstein}:
\begin{equation}
\label{eqn:polygonArea}
\beta_\text{face} = \frac{1}{2A_\text{face}} \sum_{i=1}^{n} (x_i y_{i+1} - x_{i+1} y_i),
\end{equation}
where we non-dimensionalise the result with the total area of the cell face, $A_\text{face}$. Note that the summation index is periodic so that $(n+1) = 1$. Cut cells in 2D are like cut cell faces in 3D so in a 2D simulation, \eqnref{eqn:polygonArea} would be used to compute the volume fraction of the cut cell.

As shown by Pember et al. \cite{Pember1995}, the unit interface normal (pointing into the solid), $\mathbf{\hat{n}}^b_{i,j,k}$, and the interface area $A^b_{i,j,k}$ for a cell $(i,j,k)$ can be computed from the face fractions as follows:
\begin{align}
\label{eqn:intNormal}
A^b_{i,j,k} \mathbf{\hat{n}}^b_{i,j,k} &= \Delta y \Delta z (\beta_{i-1/2,j,k} - \beta_{i+1/2,j,k}) \mathbf{\hat{i}} \nonumber \\
&+ \Delta x \Delta z (\beta_{i,j-1/2,k} - \beta_{i,j+1/2,k}) \mathbf{\hat{j}} \\
&+ \Delta x \Delta y (\beta_{i,j,k-1/2} - \beta_{i,j,k+1/2}) \mathbf{\hat{k}}. \nonumber
\end{align}
It may be noted that \eqnref{eqn:intNormal} reduces naturally to 2D and 1D.

The volume fraction of the 3D cell can be computed from the formula of the volume of a general polyhedron \cite{Goldman1991}:
\begin{equation}
\label{eqn:polyhedronVolume}
\alpha = \frac{1}{3 V_\text{cell}} \left| \sum_{i=1}^{N_F} (\mathbf{\bar{x}_i} \cdot \mathbf{\hat{n}_i}) A_i \right|,
\end{equation}
where $N_F$ is the number of faces of the polyhedron, $\mathbf{\bar{x}_i}$ is any point on the respective plane, and $\mathbf{\hat{n}_i}$ and $A_i$ are the outward unit normal and area of the plane respectively. Note that we non-dimensionalise the result by the total volume of the cell, $V_\text{cell}$.

The mesh generation procedure as described is only valid when the intersection of the cell and geometry can be described by a single interface. Sufficient resolution must therefore be used to ensure that all cut cells are `singly cut'. Note that the signed distance function can be used to deal with `split cells' created by multiple intersections of the geometry with the cell by using a `Marching Cubes' \cite{lorensen1987} based approach as in Gunther et al. \cite{gunther2011}, but this is beyond the scope of this work.

\section{Numerical method}

\label{sect:Numerical_method}

In this section, we describe the derivation of the one-dimensional KBN and LPFS fluxes, as well as their multi-dimensional extensions. A convergence and stability analysis of LPFS is presented in \sectionref{sect:Convergence_and_stability_analysis}.

\subsection{The KBN flux}

\label{sect:The_KBN_flux}

\begin{figure}
\centering
\input{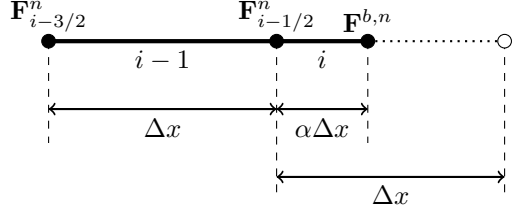}
\caption{Illustration of the KBN flux stabilisation procedure for a boundary cut cell neighbouring a regular cell in 1D.}
\label{fig:KleinStabilisation}
\end{figure}

Consider \figref{fig:KleinStabilisation}, which shows a boundary cut cell neighbouring a regular cell in 1D. Let $\mathbf{U}_{i}^{n}$ represent the conserved variable state vector for cell $i$ at time level $n$, and let $\mathbf{F}_{i \pm i/2}^{n}$ represent the explicit numerical fluxes computed at its ends.

A creative reasoning is used to compute the stabilised cut cell flux, $\mathbf{F}_{i-1/2}^{\text{KBN},n}$. In an explicit finite volume scheme, the state at the new time level is computed from the intercell fluxes at the old time level. On the other hand, if the state at the new time level were known, one could work out the stable intercell flux. An estimate of the new state, $\mathbf{\bar{U}}_{i}^{n+1}$, is calculated by extending the `influence' of the cut cell to the regular cell length (this is illustrated by the dotted line in \figref{fig:KleinStabilisation}):
\begin{equation}
\label{eqn:KleinNewStateEstimate}
\mathbf{\bar{U}}_{i}^{n+1} = \mathbf{U}_{i}^{n} + \frac{\Delta t}{\Delta x} (\mathbf{F}_{i-i/2}^{n} - \mathbf{F}^{b,n}).
\end{equation}

The actual 1D conservative update is:
\begin{equation}
\label{eqn:KleinConsUpdate}
\mathbf{\hat{U}}_{i}^{n+1} = \mathbf{U}_{i}^{n} + \frac{\Delta t}{\alpha \Delta x} (\mathbf{F}_{i-1/2}^{\text{KBN},n} - \mathbf{F}^{b,n}).
\end{equation}

Equating the right hand sides of \eqnref{eqn:KleinNewStateEstimate} and \eqnref{eqn:KleinConsUpdate} provides the expression for the stabilised flux:
\begin{equation}
\label{eqn:KleinStabilisedFlux}
\mathbf{F}_{i-1/2}^{\text{KBN},n} = \mathbf{F}^{b,n} + \alpha (\mathbf{F}_{i-i/2}^{n} - \mathbf{F}^{b,n}),
\end{equation}
which is used to update the cut cell and its neighbour (making the scheme conservative) at the regular time step $\Delta t$. Note that \eqnref{eqn:KleinStabilisedFlux} is consistent with respect to the natural limits of the grid so that as $\alpha \to 1$, $\mathbf{F}_{i-1/2}^{\text{KBN},n} \to \mathbf{F}_{i-i/2}^{n}$, and as $\alpha \to 0$, $\mathbf{F}_{i-1/2}^{\text{KBN},n} \to \mathbf{F}^{b,n}$.

\subsection{The LPFS flux}

\label{sect:The_LPFS_flux}

\begin{figure}
\centering
\input{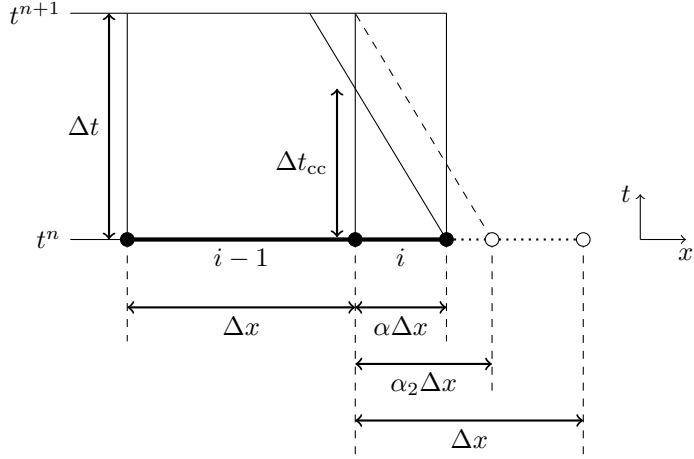}
\caption{Illustration of the LPFS flux stabilisation procedure for a boundary cut cell neighbouring a regular cell in 1D.}
\label{fig:LPFSStabilisation}
\end{figure}

From \eqnref{eqn:KleinStabilisedFlux}, it may be seen that the KBN flux uses the geometric parameter $\alpha$ to determine a stabilised flux. Here, we describe the use of geometric and wave speed information to define a new stabilised cut cell flux.

Consider \figref{fig:LPFSStabilisation}, which shows a boundary cut cell neighbouring a regular cell in the $x$-$t$ plane for one time step. $\Delta t$ is the global stable time step which is determined in part by the fastest wave speed in the domain, $W_\text{max}$ (see \eqnref{eqn:cfl}). Intuitively, the CFL time step restriction requires that no wave from the solution of the local Riemann problems should travel more than one cell width during the time step. For the configuration of \figref{fig:LPFSStabilisation}, we illustrate the `small cell problem' at the cut cell as being caused by the left-going wave from the solution of the boundary Riemann problem. Stability would therefore require the use of the smaller $\Delta t_\text{cc}$:
\begin{equation}
\label{eqn:DeltaTCC}
\Delta t_\text{cc} = C_\text{cfl} \frac{\alpha \Delta x}{W_{i}},
\end{equation}
where $W_i$ is the wave speed for the cut cell.

In the LPFS approach, we use the explicit flux $\mathbf{F}_{i-i/2}^{n}$ for the part of the time step for which it is stable, $\Delta t_\text{cc}$, and employ a different flux which can maintain stability for the duration $(\Delta t_\text{cc}, \Delta t]$. This gives the LPFS method an inherent advantage over the KBN method in regions of low velocity. Although $\alpha < 1$ depresses $\Delta t_\text{cc}$ relative to $\Delta t$, part of the reduction is offset if $W_{i} < W_\text{max}$, an effect which is most pronounced in regions of low velocity near a stagnation point. For larger cut cells, we found the ratio $\Delta t_\text{cc} / \Delta t$ to sometimes be greater than 1. The flux in this case requires no stabilisation which we allow for in LPFS by requiring that $\Delta t_\text{cc} / \Delta t \le 1$. Although we only deal with inviscid flows in this paper, it may be noted that in viscous flows, low velocity regions next to the solid boundary occur also due to the boundary layer.

There is room for creativity in the definition of the flux used in the interval $(\Delta t_\text{cc}, \Delta t]$, and we define a modified version of the original KBN flux to be used there. As described in \sectionref{sect:The_KBN_flux}, the KBN flux is derived by extending the influence of the cut cell to the full cell length. We interpret this idea in the picture of \figref{fig:LPFSStabilisation} and propose extending the influence of the cell only as far as necessary so that no wave-interface intersections occur from the solutions of the cut cell Riemann problems. This means extending the influence to a length of $\alpha_2 \Delta x$ instead of $\Delta x$ and results in the definition of a modified KBN flux:
\begin{equation}
\label{eqn:KleinModStabilisedFlux}
\mathbf{F}_{i-1/2}^{\text{KBN,mod},n} = \mathbf{F}^{b,n} + \frac{\alpha}{\alpha_2} (\mathbf{F}_{i-i/2}^{n} - \mathbf{F}^{b,n}).
\end{equation}
Like $\mathbf{F}_{i-1/2}^{\text{KBN},n}$, $\mathbf{F}_{i-1/2}^{\text{KBN,mod},n}$ is still consistent with the natural limits of the grid as $\alpha \to 1$ and $\alpha \to 0$. From \figref{fig:LPFSStabilisation}, it is also evident that $\alpha / \alpha_2$ is just $\Delta t_\text{cc} / \Delta t$.

The LPFS stabilised flux is therefore the following:
\begin{equation}
\label{eqn:LPFSStabilisedFlux}
\mathbf{F}_{i-1/2}^{\text{LPFS},n} = \frac{\Delta t_\text{cc}}{\Delta t} \mathbf{F}_{i-1/2}^{n} + \left( 1-\frac{\Delta t_\text{cc}}{\Delta t}\right) \mathbf{F}_{i-1/2}^{\text{KBN,mod},n}.
\end{equation}

Combining \eqnref{eqn:DeltaTCC} and \eqnref{eqn:cfl} gives the expression for $\Delta t_\text{cc}/\Delta t$:
\begin{equation}
\label{eqn:timeStepRatio}
\frac{\Delta t_\text{cc}}{\Delta t} = \epsilon \frac{\alpha W_\text{max}}{W_i}.
\end{equation}
For the non-linear Euler equations, we employ the `wave speeds uncertainty' parameter $\epsilon \in [0,1]$ to account for any errors arising from the use of \eqnref{eqn:wavespeeds} to estimate the cut cell wave speeds. We found setting $\epsilon$ to 0.5 to be a robust choice for the wide range of problems tackled in this paper. For the linear advection equation, $\epsilon$ is of course set precisely to 1 and, in fact, $\Delta t_\text{cc}/\Delta t = \alpha$.

Like the KBN flux, the LPFS flux is still only first order accurate at the boundary and for the simulations in this paper, we do not even linearly reconstruct the solution in the cut cells. However, the numerical results indicate that the LPFS flux not only allows the computation of more accurate solutions near stagnation points as would be expected, but also alleviates the problem of oscillatory boundary solutions produced by the KBN flux at higher Courant numbers.

\subsection{Multi-dimensional extension}

\label{sect:Multidimensional_extension}

\begin{figure}
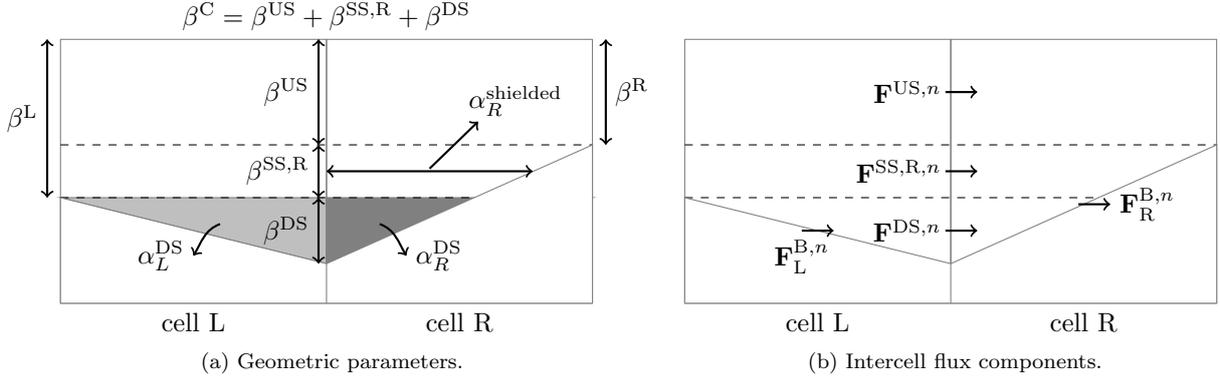

\centering
\subfloat[Geometric parameters.]{
\input{./CCStabilisationGeomParams.tfig}
\label{fig:CCStabilisationGeomParams}}
\subfloat[Intercell flux components.]{
\input{./CCStabilisationFluxes.tfig}
\label{fig:CCStabilisationFluxes}}
\caption{Illustration of the parameters used in the flux stabilisation process for multi-dimensional simulations.}
\label{fig:CCStabilisationParams}
\end{figure}

The LPFS flux can be implemented for multi-dimensional simulations in the framework of Klein et al. which requires attention to be given to the irregular nature of the cut cells. With reference to \figref{fig:CCStabilisationParams}, we explain the flux calculation procedure at the interface $I^\text{face}_\text{C}$ between two neighbouring cells for an arbitrary dimensional sweep in a 2D simulation. The procedure is exactly the same for 3D simulations.

The interface is divided into four regions with respect to the current sweep direction:
\begin{itemize}[-]
\item The `unshielded' (US) region which does not `face' a boundary in the current coordinate direction.
\item The `singly-shielded from the right' (SS,R) region which is `covered' by the boundary from the right.
\item The `singly-shielded from the left' (SS,L) region which is covered by the boundary from the left.
\item The `doubly-shielded' (DS) region which faces the boundary from the left and right.
\end{itemize}

In a singly-shielded region, $\alpha^\text{shielded}_K$ (where $K$=$L$, $R$ as appropriate) represents the average distance from $I^\text{face}_\text{C}$ to the boundary in the current coordinate direction, non-dimensionalised by the corresponding regular cell spacing. $\alpha^\text{DS}_L$ and $\alpha^\text{DS}_R$ are the fluid volume fractions in the doubly-shielded regions of the left and right cells respectively. All the parameters in \figref{fig:CCStabilisationGeomParams} can be computed from the geometric information described in \sectionref{sect:Cut_cell_mesh_generation}. The details are left out for the sake of brevity.

The fluxes labelled in \figref{fig:CCStabilisationFluxes} are calculated as follows:
\begin{itemize}[-]
\item The boundary flux, for example, $\mathbf{F}^{\text{B},n}_\text{L}$, is calculated by evaluating the flux acting in the current coordinate direction for the `boundary state' given by the solution of the wall normal Riemann problem. Note that in order to maintain conservation in a dimensionally split framework, the advective boundary fluxes need to be computed using `reference' boundary states computed at the start of a time step and kept constant in between sweeps. This restriction does not apply to other fluxes, so that for the Euler equations \eqnref{eqn:eul_eqns}, for example, the boundary state pressure can, in fact, be updated in each sweep. A specification of this non-obvious requirement for computing the boundary fluxes was one of the unique insights provided by Klein et al. \cite{Klein2009}.
\item $\mathbf{F}^{\text{US},n}$ is taken to be the standard explicit 1D flux for regular cells since it is not shielded by the boundary.
\item The singly-shielded fluxes, $\mathbf{F}^{\text{SS,L},n}$ and $\mathbf{F}^{\text{SS,R},n}$, are calculated as one-dimensional LPFS stabilised fluxes as per \eqnref{eqn:LPFSStabilisedFlux}, with the place of $\alpha$ in \eqnref{eqn:timeStepRatio} being taken by $\alpha^\text{shielded}_L$ or $\alpha^\text{shielded}_R$ respectively.
\item In the doubly-shielded region, there is a genuine restriction on the time step imposed by the distance between the two boundaries. Here, we propose the use of a simple conservative `mixing' flux designed to produce the same volume-averaged solution in the doubly-shielded parts of the cells over the course of the dimensional sweep:
\begin{equation}
\label{eqn:MixingFlux}
\mathbf{F}^{\text{DS},n} = \frac{1}{(\alpha^\text{DS}_L + \alpha^\text{DS}_R)} \left[ \frac{\alpha^\text{DS}_L \alpha^\text{DS}_R \Delta x}{\beta^\text{DS} \Delta t} (\mathbf{U}_L^n - \mathbf{U}_R^n) + \alpha^\text{DS}_L \mathbf{F}^{\text{B},n}_\text{R} + \alpha^\text{DS}_R \mathbf{F}^{\text{B},n}_\text{L} \right].
\end{equation}
\end{itemize}

The modified flux $\mathbf{F}^{\text{modified},n}_\text{C}$ at $I^\text{face}_\text{C}$ is taken as an area-weighted sum of the individual components:
\begin{equation}
\label{eqn:MultiDimensionalFlux}
\mathbf{F}^{\text{modified},n}_\text{C} = \frac{1}{\beta^\text{C}} \left[ \beta^\text{US} \mathbf{F}^{\text{US},n} + \beta^\text{SS,L} \mathbf{F}^{\text{SS,L},n} + \beta^\text{SS,R} \mathbf{F}^{\text{SS,R},n} + \beta^\text{DS} \mathbf{F}^{\text{DS},n} \right].
\end{equation}

The multi-dimensional update formula for a cut cell of index $(i,j)$ in 2D using Godunov splitting would be:
\begin{align}
\label{eqn:MultiDimensionalUpdate}
&\mathbf{U}_{i,j}^{n+1/2} = \mathbf{U}_{i,j}^{n} + \frac{\Delta t}{\alpha_{i,j} \Delta x} \left[ \beta_{i-1/2,j} \mathbf{F}_{i-1/2,j}^{\text{modified},n} - \beta_{i+1/2,j} \mathbf{F}_{i+1/2,j}^{\text{modified},n} - \left( \beta_{i-1/2,j} - \beta_{i+1/2,j} \right) \mathbf{F}_{i,j}^{\text{B},n} \right],\nonumber\\
&\mathbf{U}_{i,j}^{n+1} = \mathbf{U}_{i,j}^{n+1/2} + \frac{\Delta t}{\alpha_{i,j} \Delta y} \left[ \beta_{i,j-1/2} \mathbf{G}_{i,j-1/2}^{\text{modified},n} - \beta_{i,j+1/2} \mathbf{G}_{i,j+1/2}^{\text{modified},n} - \left( \beta_{i,j-1/2} - \beta_{i,j+1/2} \right) \mathbf{G}_{i,j}^{\text{B},n} \right],
\end{align}
where we use $\mathbf{F}$ and $\mathbf{G}$ to denote fluxes acting in the $x$ and $y$ directions respectively. A 3D simulation would involve one more sweep using fluxes $\mathbf{H}$ acting in the $z$ direction.

\subsubsection{Post-sweep correction at concavities}

\label{sect:Post_sweep_correction_at_concavities}

When $\beta^L = \beta^R = 0$, i.e., when $I^\text{face}_\text{C}$ is completely shielded by the boundary from both sides, we found that the mixing flux \eqnref{eqn:MixingFlux} was sometimes unable to maintain stability. A simple conservative fix for this problem is to merge the solution in such pairs of `fully doubly-shielded' cells with that of their immediate neighbours in a volume-fraction weighted manner at the end of the sweep. This is not computationally expensive as the affected cells can be identified at the flux computation stage and directly targeted after the sweep. The performance of this strategy is demonstrated through the results of \sectionref{sect:Shock_reflection_over_a_double_wedge} and \sectionref{sect:Space_reentry_vehicle_simulation}, which contain examples of fully doubly-shielded cells in two and three dimensions respectively.

The design of a better doubly-shielded flux $\mathbf{F}^{\text{DS},n}$ which would avoid the need for this post-sweep correction is identified as an open research problem. It should be noted, however, that the correction even as it stands affects only a small number of cells. For the complex geometry test case of \sectionref{sect:Space_reentry_vehicle_simulation}, for example, `fully doubly-shielded' cells make up less than $0.05\%$ of all cut cells in a given dimensional sweep.

\section{Convergence and stability analysis}

\label{sect:Convergence_and_stability_analysis}

\begin{figure}
\centering
\input{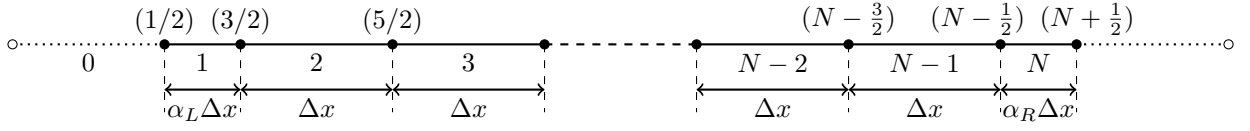}
\caption{1D mesh with cut cells located at the left and right edges of the domain.}
\label{fig:1DCCMesh}
\end{figure}

In this section, we demonstrate the convergence and stability of the first order LPFS scheme for the solution of the linear advection equation
\begin{equation}
\label{eqn:linadv}
u_{t} + au_{x} = 0,
\end{equation}
on the one-dimensional mesh consisting of $N$ cells shown in \figref{fig:1DCCMesh}. The boundary cells `$1$' and `$N$' are both assumed to be cut cells with respective volume fractions $\alpha_L$ and $\alpha_R$. For the purposes of the analysis, we can therefore assume without loss of generality that $a > 0$.

From \eqnref{eqn:LPFSStabilisedFlux}, the stabilised LPFS flux at interface ($3/2$) can be worked out to be:
\begin{align}
		f_{3/2}^{\text{LPFS},n} &= \alpha_L f_{3/2}^{n} + (1-\alpha_L)f_{3/2}^{\text{KBN},n} \nonumber \\
		                    &= \alpha_L f_{3/2}^{n} + (1-\alpha_L)[f_{1/2}^{n} + \alpha_L(f_{3/2}^{n} - f_{1/2}^{n})] \nonumber \\
		                    &= (\alpha_L-1)^2 f_{1/2}^{n} +\alpha_L (2-\alpha_L)f_{3/2}^{n}.
\end{align}

Similarly, the stabilised LPFS flux at interface ($N-1/2$) is:
\begin{align}
		f_{N-1/2}^{\text{LPFS},n} &= \alpha_R f_{N-1/2}^{n} + (1-\alpha_R)f_{N-1/2}^{\text{KBN},n} \nonumber \\
		                    &= \alpha_R f_{N-1/2}^{n} + (1-\alpha_R)[f_{N+1/2}^{n} + \alpha_R(f_{N-1/2}^{n} - f_{N+1/2}^{n})] \nonumber \\
		                    &= \alpha_R(2-\alpha_R) f_{N-1/2}^{n} + (\alpha_R-1)^2 f_{N+1/2}^{n}.
\end{align}

The update formulas for cells `$1$', `$2$', `$N-1$' and `$N$' which are affected by the flux stabilisation are thus:
\begin{align}
		U_1^{n+1} &= U_1^{n} + c(2-\alpha_L)(U_0^n - U_1^n), \label{eqn:cell1Update} \\
		U_2^{n+1} &= U_2^{n} + c[(\alpha_L-1)^2 U_0^n +\alpha_L (2-\alpha_L)U_1^n - U_2^n], \\
		U_{N-1}^{n+1} &= U_{N-1}^{n} + c[U_{N-2}^n - \alpha_R(2-\alpha_R)U_{N-1}^n - (\alpha_R-1)^2 U_{N}^n], \\
		U_N^{n+1} &= U_N^{n} + c(2-\alpha_R)(U_{N-1}^n-U_{N}^n), \label{eqn:cellNUpdate}
\end{align}
where $c=\frac{a \Delta t}{\Delta x}$ is the Courant number. Note that we use capital $U_i^n$ to denote the discrete approximation of the volume averaged solution in cell $i$ at time $n$.

\subsection{`Supraconvergence' property of the LPFS scheme}

\label{sect:Supraconvergence_property_of_the_LPFS_scheme}

Attempting a truncation error analysis of the scheme in any of the cells `$1$', `$2$', `$N-1$' and `$N$' affected by the flux stabilisation shows an inconsistency with the governing equation \eqnref{eqn:linadv}. Note that in the following, we use small $u_i^n$ to represent the grid function of the exact solution $u(x,t)$ in cell $i$ at time $n$.

Consider, for example, cell $N$. The truncation error in that cell, $Lu_N$, can be found to be:
\begin{align*}
Lu_N &= \frac{u_N^{n+1}-u_N^{n}}{\Delta t} -(2-\alpha_R)a\frac{u_{N-1}^n - u_N^n}{\Delta x} \\
     &= u_t(x_N,t^n) -(2-\alpha_R)a\frac{u_N^n - \frac{(1+\alpha_R)}{2}\Delta x u_x(x_N,t^n) - u_{N}^n}{\Delta x} + \mathcal{O}(\Delta t, \Delta x) \\
     &= -a u_x(x_N,t^n) + (2-\alpha_R)a\frac{(1+\alpha_R)}{2} u_x(x_N,t^n) + \mathcal{O}(\Delta t, \Delta x) \\
     &= \frac{1}{2}\alpha_R (\alpha_R-1) a u_x(x_N,t^n) + \mathcal{O}(\Delta t, \Delta x),
\end{align*}
so that the scheme is inconsistent unless $\alpha_R$ is 1. Therefore, we cannot invoke the `Lax equivalence theorem' \cite{lax1956} in the analysis.

Despite this inconsistency, numerical tests show that the scheme does, in fact, converge with first order accuracy (see \sectionref{sect:Convergence_tests}). To prove this `supraconvergence' property of the method, we follow the approach of Berger et al. \cite{berger2003h} who perform a similar analysis for their $h$-box method.

The aim is to find another grid function $w$ which differs from the grid function of $u$ by an $\mathcal{O}(\Delta x)$ amount, and for which the truncation error in all cells is $\mathcal{O}(\Delta t, \Delta x)$. This gives
\begin{equation}
|w_i-U_i| = \mathcal{O}(\Delta t, \Delta x)
\end{equation}
for $i = 1,\dots,N$. However, since $w = u + \mathcal{O}(\Delta x)$, this would imply that
\begin{equation}
|u_i-U_i| = \mathcal{O}(\Delta t, \Delta x)
\end{equation}
for $i = 1,\dots,N$, i.e. that the scheme does in fact approximate the true solution to first order despite the inconsistency in the boundary cells and their immediate neighbours.

Let the grid function $w$ be defined as follows:
\begin{equation}
\label{eqn:wDefinition}
w_i^n =
\begin{cases}
    u_i^n + \frac{\alpha_L(\alpha_L - 1)}{2(2-\alpha_L)}\Delta x u_x(x_i,t^n) &\text{ if } i=1,\\
    u_{i}^{n} + \left( \frac{1+\alpha_R(\alpha_R-2)}{\alpha_R-2} \right)\Delta x u_x(x_{i},t^n) &\text{ if } i=N-1,\\
    u_{i}^{n} + \frac{1}{2} (\alpha_R-1)\Delta x u_x(x_{i},t^n) &\text{ if } i=N,\\
    u_i^n &\text{ otherwise},
\end{cases}
\end{equation}
where we note that $w = u + \mathcal{O}(\Delta x)$ as desired.

The truncation error of $w$ in cell `$1$', $Lw_1$ can be worked out to be:
\begin{align*}
Lw_1 &= \frac{w_1^{n+1}-w_1^{n}}{\Delta t} -(2-\alpha_L)a\frac{w_{0}^n - w_1^n}{\Delta x} \\
     &= \frac{u_1^{n+1} + \frac{\alpha_L(\alpha_L - 1)}{2(2-\alpha_L)}\Delta x u_x(x_1,t^{n+1}) - u_1^{n} - \frac{\alpha_L(\alpha_L - 1)}{2(2-\alpha_L)}\Delta x u_x(x_1,t^{n})}{\Delta t} \\
     &-(2-\alpha_L)a\frac{u_{0}^n - u_1^{n} - \frac{\alpha_L(\alpha_L - 1)}{2(2-\alpha_L)}\Delta x u_x(x_1,t^{n})}{\Delta x} \\
     &= \frac{u_1^{n} + \Delta t u_t(x_1,t^n) + \frac{\alpha_L(\alpha_L - 1)}{2(2-\alpha_L)}\Delta x u_x(x_1,t^{n}) - u_1^{n} - \frac{\alpha_L(\alpha_L - 1)}{2(2-\alpha_L)}\Delta x u_x(x_1,t^{n})}{\Delta t} \\
     &-(2-\alpha_L)a\frac{u_{1}^n - \frac{(1+\alpha_L)}{2}\Delta x u_x(x_1,t^{n}) - u_1^{n} - \frac{\alpha_L(\alpha_L - 1)}{2(2-\alpha_L)}\Delta x u_x(x_1,t^{n})}{\Delta x} + \mathcal{O}(\Delta t, \Delta x) \\
     &= u_t(x_1,t^n) -(2-\alpha_L)a\frac{-\frac{\Delta x u_x(x_1,t^{n})}{(2-\alpha_L)}}{\Delta x} + \mathcal{O}(\Delta t, \Delta x) \\
     &= u_t(x_1,t^n) + au_x(x_1,t^{n}) + \mathcal{O}(\Delta t, \Delta x) = \mathcal{O}(\Delta t, \Delta x).
\end{align*}

Similarly, the truncation error in cell `$2$', $Lw_2$ is:
\begin{align*}
Lw_2 &= \frac{w_2^{n+1}-w_2^{n}}{\Delta t} -a\frac{[(\alpha_L-1)^2 w_0^n + \alpha_L(2-\alpha_L)w_1^n - w_2^n]}{\Delta x} \\
     &= \frac{u_2^{n+1}-u_2^{n}}{\Delta t} - a\frac{\left[(\alpha_L-1)^2 u_0^n + \alpha_L(2-\alpha_L)\left(u_1^n + \frac{\alpha_L(\alpha_L - 1)}{2(2-\alpha_L)}\Delta x u_x(x_1,t^n)\right)- u_2^n \right]}{\Delta x} \\
     &= u_t(x_2,t^n) - a\frac{\left[ (\alpha_L-1)^2 (u_2^n - (1+\alpha_L)\Delta x u_x(x_2,t^{n}))\right]}{\Delta x} \\
     &- a\frac{\left[\alpha_L(2-\alpha_L) \left(u_2^n - \frac{(1+\alpha_L)}{2}\Delta x u_x(x_2,t^n) + \frac{\alpha_L(\alpha_L - 1)}{2(2-\alpha_L)}\Delta x u_x(x_2,t^n)\right) - u_2^n\right]}{\Delta x} + \mathcal{O}(\Delta t, \Delta x) \\
     &= u_t(x_2,t^n) + au_x(x_2,t^{n}) + \mathcal{O}(\Delta t, \Delta x) = \mathcal{O}(\Delta t, \Delta x).
\end{align*}

At the right edge of the domain, the truncation error in cell `$N-1$', $Lw_{N-1}$ is:
\begin{align*}
Lw_{N-1} &= \frac{w_{N-1}^{n+1}-w_{N-1}^{n}}{\Delta t} -a\frac{[w_{N-2}^n - \alpha_R(2-\alpha_L)w_{N-1}^n - (\alpha_R-1)^2 w_{N}^n]}{\Delta x} \\
         &= \frac{u_{N-1}^{n+1} + \left( \frac{1+\alpha_R(\alpha_R-2)}{\alpha_R-2} \right)\Delta x u_x(x_{N-1},t^{n+1}) - u_{N-1}^{n} - \left( \frac{1+\alpha_R(\alpha_R-2)}{\alpha_R-2} \right)\Delta x u_x(x_{N-1},t^{n})}{\Delta t} \\
         &-a\frac{\left[u_{N-2}^n - \alpha_R(2-\alpha_L) \left(u_{N-1}^{n} + \left( \frac{1+\alpha_R(\alpha_R-2)}{\alpha_R-2} \right)\Delta x u_x(x_{N-1},t^n)\right)\right]}{\Delta x} \\
         &-a\frac{\left[ - (\alpha_R-1)^2 \left( u_{N}^{n} + \frac{1}{2} (\alpha_R-1)\Delta x u_x(x_{N},t^n) \right) \right]}{\Delta x} \\
         &= \frac{u_{N-1}^{n+1} + \left( \frac{1+\alpha_R(\alpha_R-2)}{\alpha_R-2} \right)\Delta x u_x(x_{N-1},t^{n}) - u_{N-1}^{n} - \left( \frac{1+\alpha_R(\alpha_R-2)}{\alpha_R-2} \right)\Delta x u_x(x_{N-1},t^{n})}{\Delta t} \\
         &-a\frac{\left[u_{N-1}^n - \Delta x u_x(x_{N-1},t^n) - \alpha_R(2-\alpha_L) \left(u_{N-1}^{n} + \left( \frac{1+\alpha_R(\alpha_R-2)}{\alpha_R-2} \right)\Delta x u_x(x_{N-1},t^n)\right)\right]}{\Delta x} \\
         &-a\frac{\left[ - (\alpha_R-1)^2 \left( u_{N-1}^{n} + \frac{(1+\alpha_R)}{2} \Delta x u_x(x_{N-1},t^n) + \frac{1}{2} (\alpha_R-1)\Delta x u_x(x_{N-1},t^n) \right) \right]}{\Delta x} + \mathcal{O}(\Delta t, \Delta x) \\
         &= u_t(x_{N-1},t^n) + au_x(x_{N-1},t^{n}) + \mathcal{O}(\Delta t, \Delta x) = \mathcal{O}(\Delta t, \Delta x).
\end{align*}

Similarly, the the truncation error in cell `$N$', $Lw_N$ is:
\begin{align*}
Lw_N &= \frac{w_N^{n+1}-w_N^{n}}{\Delta t} -(2-\alpha_R)a\frac{w_{N-1}^n - w_N^n}{\Delta x} \\
     &= \frac{u_{N}^{n+1} + \frac{1}{2} (\alpha_R-1)\Delta x u_x(x_{N},t^{n+1}) - u_{N}^{n} - \frac{1}{2} (\alpha_R-1)\Delta x u_x(x_{N},t^{n})}{\Delta t} \\
     &-(2-\alpha_R)a\frac{u_{N-1}^{n} + \left( \frac{1+\alpha_R(\alpha_R-2)}{\alpha_R-2} \right)\Delta x u_x(x_{N-1},t^n) - u_{N}^{n} - \frac{1}{2} (\alpha_R-1)\Delta x u_x(x_{N},t^{n})}{\Delta x} \\
     &= \frac{u_{N}^{n+1} + \frac{1}{2} (\alpha_R-1)\Delta x u_x(x_{N},t^{n}) - u_{N}^{n} - \frac{1}{2} (\alpha_R-1)\Delta x u_x(x_{N},t^{n})}{\Delta t} \\
     &-(2-\alpha_R)a\frac{u_{N}^{n} - \frac{(1+\alpha_R)}{2}\Delta x u_x(x_{N},t^{n}) + \left( \frac{1+\alpha_R(\alpha_R-2)}{\alpha_R-2} \right)\Delta x u_x(x_{N},t^n)}{\Delta x} \\
     &-(2-\alpha_R)a\frac{- u_{N}^{n} - \frac{1}{2} (\alpha_R-1)\Delta x u_x(x_{N},t^{n})}{\Delta x} + \mathcal{O}(\Delta t, \Delta x) \\
     &= u_t(x_{N},t^n) -(2-\alpha_R)a\frac{-\frac{\Delta x u_x(x_N,t^{n})}{(2-\alpha_R)}}{\Delta x} + \mathcal{O}(\Delta t, \Delta x) \\
     &= u_t(x_{N},t^n) + au_x(x_{N},t^n) + \mathcal{O}(\Delta t, \Delta x) = \mathcal{O}(\Delta t, \Delta x).
\end{align*}

For all other cells which are updated with the regular upwind formula, since we define $w_i^n$ to be equal to $u_i^n$, the truncation error is again $\mathcal{O}(\Delta t, \Delta x)$ as desired, which concludes the proof.

\subsection{Stability of the LPFS scheme}

\label{sect:Stability_of_the_LPFS_scheme}

In the previous \sectionref{sect:Supraconvergence_property_of_the_LPFS_scheme}, we have shown that the computed $U_i^n$ approximate the $w_i^n$ (and hence, the $u_i^n$) to first order assuming that the scheme is stable. In this section, we prove that the $U_i^n$ approach the $w_i^n$ in a stable manner for $c \in (0,1]$ for all $i = 1,\dots,N$.

Consider the error function $v_i^n = U_i^n - w_i^n$. Given some sufficiently smooth initial conditions, $v_i^0$ is $\mathcal{O}(\Delta t, \Delta x)$ for all cells. For the cells unaffected by the flux stabilisation, stability is already guaranteed for $c \in (0,1]$ since they are updated using the regular upwind formula. $v_i^n$ in those cells continues to remain first order as $n$ increases.

As before, we therefore need only focus our attention on cells `$1$', `$2$', `$N-1$' and `$N$' and investigate how $v_i^n$ evolves for those cells as $n \to \infty$. Recall from the supraconvergence analysis of \sectionref{sect:Supraconvergence_property_of_the_LPFS_scheme} that:
\begin{align}
		w_1^{n+1} &= w_1^{n} + c(2-\alpha_L)(w_0^n - w_1^n) + \mathcal{O}(\Delta t, \Delta x), \label{eqn:cell1wUpdatev2} \\
		w_2^{n+1} &= w_2^{n} + c[(\alpha_L-1)^2 w_0^n +\alpha_L (2-\alpha_L)w_1^n - w_2^n] + \mathcal{O}(\Delta t, \Delta x), \\
		w_{N-1}^{n+1} &= w_{N-1}^{n} + c[w_{N-2}^n - \alpha_R(2-\alpha_R)w_{N-1}^n - (\alpha_R-1)^2 w_{N}^n] + \mathcal{O}(\Delta t, \Delta x), \\
		w_N^{n+1} &= w_N^{n} + c(2-\alpha_R)(w_{N-1}^n-w_{N}^n) + \mathcal{O}(\Delta t, \Delta x). \label{eqn:cellNwUpdatev2}
\end{align}

Using \eqnref{eqn:cell1Update}-\eqnref{eqn:cellNUpdate} and \eqnref{eqn:cell1wUpdatev2}-\eqnref{eqn:cellNwUpdatev2}, it is straightforward to work out $v_1^n$, $v_2^n$, $v_{N-1}^n$ and $v_N^n$. At the left edge of the domain,
\begin{align}
	v_1^{n+1} &= v_1^{n} + c(2-\alpha_L)(v_0^n - v_1^n) + \mathcal{O}(\Delta t, \Delta x), \nonumber \\
						&= v_1^{n} + c(2-\alpha_L)(- v_1^n) + \mathcal{O}(\Delta t, \Delta x) \label{eqn:v1},
\end{align}
since $v_0^{n} = \mathcal{O}(\Delta t,\Delta x)$ by definition (\eqnref{eqn:wDefinition}). Similarly,
\begin{align}
	v_2^{n+1} &= v_2^{n} + c[(\alpha_L-1)^2 v_0^n +\alpha_L (2-\alpha_L)v_1^n - v_2^n] + \mathcal{O}(\Delta t, \Delta x), \nonumber \\
						&= v_2^{n} + c[\alpha_L (2-\alpha_L)v_1^n - v_2^n] + \mathcal{O}(\Delta t, \Delta x) \label{eqn:v2}.
\end{align}

At the right edge of the domain,
\begin{align}
	v_{N-1}^{n+1} &= v_{N-1}^{n} + c[v_{N-2}^n - \alpha_R(2-\alpha_R)v_{N-1}^n - (\alpha_R-1)^2 v_{N}^n] + \mathcal{O}(\Delta t, \Delta x), \nonumber \\
								&= v_{N-1}^{n} + c[- \alpha_R(2-\alpha_R)v_{N-1}^n - (\alpha_R-1)^2 v_{N}^n] + \mathcal{O}(\Delta t, \Delta x) \label{eqn:vNm1},
\end{align}
and
\begin{align}
	v_N^{n+1} &= v_N^{n} + c(2-\alpha_R)(v_{N-1}^n-v_{N}^n) + \mathcal{O}(\Delta t, \Delta x) \label{eqn:vN}.
\end{align}

We can now use \eqnref{eqn:v1}, \eqnref{eqn:v2}, \eqnref{eqn:vNm1} and \eqnref{eqn:vN} to write the following linear inhomogeneous recurrence relation for $\mathbf{x}^{n+1} = [v_{1}^{n+1} \:\: v_{2}^{n+1} \:\: v_{N-1}^{n+1} \:\: v_{N}^{n+1}]^T$:
\begin{equation}
\label{eqn:recurrenceLPFS}
\begin{split}
\begin{sbmatrix}{\mathbf{x}^{n+1}}
  v_{1}^{n+1} \\[0.7em]
  v_{2}^{n+1} \\[0.7em]
  v_{N-1}^{n+1} \\[0.7em]
  v_{N}^{n+1}
\end{sbmatrix}
=
\begin{sbmatrix}{A}
  1-c(2-\alpha_L) & 0 & 0 & 0 \\[0.7em]
  c\alpha_L(2-\alpha_L) & 1-c & 0 & 0 \\[0.7em]
  0 & 0 & 1-c\alpha_R (2-\alpha_R) & -c(\alpha_R-1)^2 \\[0.7em]
  0 & 0 & c(2-\alpha_R) & 1-c(2-\alpha_R)
\end{sbmatrix}
\begin{sbmatrix}{\mathbf{x}^{n}}
  v_{1}^{n} \\[0.7em]
  v_{2}^{n} \\[0.7em]
  v_{N-1}^{n} \\[0.7em]
  v_{N}^{n}
\end{sbmatrix}
+
\begin{sbmatrix}{\mathbf{b}^{n}}
  \mathcal{O}(\Delta t,\Delta x) \\[0.7em]
  \mathcal{O}(\Delta t,\Delta x) \\[0.7em]
  \mathcal{O}(\Delta t,\Delta x) \\[0.7em]
  \mathcal{O}(\Delta t,\Delta x)
\end{sbmatrix}.
\end{split}
\end{equation}

We therefore have
\begin{align*}
\mathbf{x}^{n+1} &= A\mathbf{x}^n + \mathbf{b}^n \\
          &= A(A\mathbf{x}^{n-1} + \mathbf{b}^{n-1}) + \mathbf{b}^n = A^2\mathbf{x}^{n-1} + A\mathbf{b}^{n-1} + \mathbf{b}^n \\
          &= A^3\mathbf{x}^{n-2} + A^2\mathbf{b}^{n-2} + A\mathbf{b}^{n-1} + \mathbf{b}^n \\
          &= A^{n+1}\mathbf{x}^{0} + \sum_{\nu=0}^{n} A^{\nu} \mathbf{b}^{n-\nu},
\end{align*}
where we note that $\mathbf{x}^{0}$ and all the $\mathbf{b}^{n-\nu}$ vectors are made up of $\mathcal{O}(\Delta t,\Delta x)$ components. Hence,
\begin{equation*}
|\mathbf{x}^{n+1}| \le \|A\|^{n+1}|\mathbf{x}^{0}| + \max_{k=0,\dots,n}\{|\mathbf{b}^k| \} \sum_{\nu=0}^{n} \|A\|	^{\nu}.
\end{equation*}

In order for $|\mathbf{x}^{n+1}|$ to remain a bounded $\mathcal{O}(\Delta t,\Delta x)$ term, it is clear that we require $\|A\| < 1$ for all $c \in (0,1]$ and $\alpha_L,\alpha_R \in (0,1]$. For the first term to the right hand side of the inequality, this would imply that $\|A\|^{n+1}|\mathbf{x}^{0}| < |\mathbf{x}^{0}|$. The second term on the right hand side, on the other hand, can be thought of as a geometric series with the `common ratio' of two consecutive terms being $\|A\|$. $\|A\| < 1$ ensures that the series converges to a finite sum as $n \to \infty$.

To study the behaviour of $\|A\|$, we make use of the convenient result that the spectral radius of $A$, $\rho(A)$, is the infimum of $\|A\|$, as $\|\cdot\|$ ranges over the set of matrix norms \cite{serre2000matrices}. If we can confirm that the eigenvalues of $A$ have magnitude less than 1 for all $c \in (0,1]$ and $\alpha_L,\alpha_R \in (0,1]$, we would have proved the stability of the recursion.

Since $A$ is a block diagonal matrix, its eigenvalues will be the union of the eigenvalues of the following $2 \times 2$ matrices
\begin{equation}
\label{eqn:ALAR}
A_{L} =
\begin{bmatrix}
  1-c(2-\alpha_L) & 0 \\[0.7em]
  c\alpha_L(2-\alpha_L) & 1-c
\end{bmatrix},
\:
A_{R} =
\begin{bmatrix}
	1-c\alpha_R (2-\alpha_R) & -c(\alpha_R-1)^2 \\[0.7em]
	c(2-\alpha_R) & 1-c(2-\alpha_R)
\end{bmatrix},
\end{equation}
which are positioned along its main diagonal.

The eigenvalues of $A_{L}$ are
\begin{align*}
	\lambda_{L}^{1} &= 1-c,\\
	\lambda_{L}^{2} &= 1-c(2-\alpha_{L}).
\end{align*}
It is easy to verify that
\begin{enumerate}
	\item $\lambda_{L}^{1}$ and $\lambda_{L}^{2}$ are real,
	\item $\lambda_{L}^{1} \in [0,1)$ for $c \in (0,1]$, and
	\item $|\lambda_{L}^{2}| < 1$ for $c \in (0,1]$ and $\alpha_L \in (0,1]$,
\end{enumerate}
so that the eigenvalues of $A_{L}$ have magnitude less than 1 over the full range of $c$ and $\alpha_{L}$ as desired.

The eigenvalues of $A_{R}$, on the other hand, are
\begin{align*}
	\lambda_{R}^{1} &= \frac{1}{2}\left[2-2c-\alpha_{R}c+\alpha_{R}^{2}c-(\alpha_{R}-1)\left(\sqrt{-4+\alpha_{R}^{2}}\right)c\right],\\
	\lambda_{R}^{2} &= \frac{1}{2}\left[2-2c-\alpha_{R}c+\alpha_{R}^{2}c+(\alpha_{R}-1)\left(\sqrt{-4+\alpha_{R}^{2}}\right)c\right].
\end{align*}
Since the $\sqrt{-4+\alpha_{R}^{2}}$ term is imaginary for all $\alpha_{R} \in (0,1]$, it is clear that $\lambda_{R}^{1}$ and $\lambda_{R}^{2}$ are complex conjugates. The square of the magnitude of any one of the eigenvalues, $|\lambda_{R}|^2$, is then
\begin{equation}
	\label{eqn:lambdaRDet}
	|\lambda_{R}|^2 = \lambda_{R}^{1}\lambda_{R}^{2} = \text{det}(A_{R}) > 0.
\end{equation}
To ensure that $|\lambda_{R}| < 1$, we therefore only need to check that $\text{det}(A_{R}) < 1$. We have
\begin{equation}
	\text{det}(A_{R}) = 1-[2c(1-c)+\alpha_{R}c(1-\alpha_{R}+c)],
\end{equation}
where it can be noted straightaway that $\text{det}(A_{R}) < 1$ since $2c(1-c) > 0$ and $\alpha_{R}c(1-\alpha_{R}+c) > 0$ for $c, \alpha_R \in (0,1]$.

\begin{figure}[H]
\centering
\includegraphics[width=0.45\textwidth]{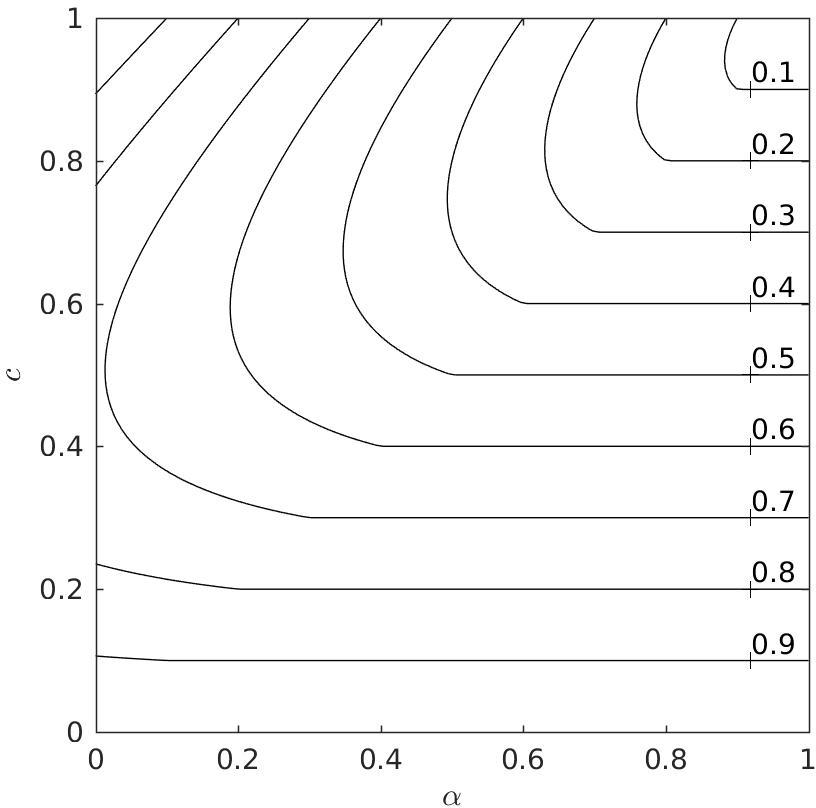}
\caption{Contour plot of the spectral radius of $A$.}
\label{fig:ASpectralRadius}
\end{figure}

This concludes the proof. We can therefore confirm that $\|A\|<1$ for the full range of $c$, $\alpha_{L}$ and $\alpha_{R}$ and that the $U_i^n$ stably approximate the $w_i^n$ (and hence, the $u_i^n$) to first order in all cells as desired. \figref{fig:ASpectralRadius} shows a contour plot of the spectral radius of $A$ for illustrative purposes.

\section{Results}

\label{sect:Results}

In this section, we present the results of computations for a number of multi-dimensional test problems for the linear advection and Euler equations. All simulations were run at $C_\text{cfl} = 0.8$.

\subsection{Convergence tests}

\label{sect:Convergence_tests}

The numerical order of convergence of LPFS is investigated through a series of advection tests. Note that for these smooth test problems, we do not use the limiter.

The $L_p$ norm of the error for a variable $\phi$ at a resolution $\Delta x$ is computed as
\begin{equation}
\label{eqn:LpNorm}
L_p^{\Delta x} = \left( \frac{1}{N} \sum_{i=1}^{N} (|\phi_i^\text{sim} - \phi_i^\text{exact}|)^{p} \right)^\frac{1}{p},
\end{equation}
where $N$ is the total number of cells, $\phi_i^\text{sim}$ is the numerical solution in cell $i$, and $\phi_i^\text{exact}$ is the exact solution evaluated at the volumetric centroid of cell $i$. The numerical order of convergence is estimated as
\begin{equation}
\label{eqn:numericalorder}
L_p \: \text{order} = \frac{\log\left(\frac{L_p^{2 \Delta x}}{L_p^{\Delta x}}\right)}{\log(2)}.
\end{equation}

With first order accuracy at the boundary but second order accuracy in regular cells, we show briefly how the computed order depends on the value of $p$. Consider that the total number of cells in the domain scales as $\mathcal{O}(n^D)$, where $n$ is the number of cells along one dimension, and $D$ is the number of dimensions, while the number of cut cells scales as $\mathcal{O}(n^{D-1})$. Further, $n$ scales as $\mathcal{O}(\Delta x^{-1})$. Substituting these values in \eqnref{eqn:LpNorm} gives:
\begin{align}
\label{eqn:totalPorder}
L_p^{\Delta x} &= \left( \frac{\mathcal{O}(\Delta x^pn^{D-1}) + \mathcal{O}(\Delta x^{2p}) [\mathcal{O}(n^D)-\mathcal{O}(n^{D-1})]}{\mathcal{O}(n^D)} \right)^\frac{1}{p} \nonumber\\
&= \left( \mathcal{O}(\Delta x^{p+1}) + \mathcal{O}(\Delta x^{2p}) - \mathcal{O}(\Delta x^{2p+1}) \right)^\frac{1}{p} \\
&= \mathcal{O}(\Delta x^{\frac{p+1}{p}}) \nonumber.
\end{align}

Hence, we would expect the $L_1$ norm to converge as $\mathcal{O}(\Delta x^2)$, the $L_2$ norm to converge as $\mathcal{O}(\Delta x^{1.5})$ and the $L_\infty$ norm to converge as $\mathcal{O}(\Delta x)$.

\subsubsection{One-dimensional advection}

This problem involves the linear advection of the smooth profile,
\begin{equation}
u(x) = \sin(2\pi x),
\end{equation}
to the right at a speed $a = 1.0$ in the interval $x \in [0,1]$ with periodic boundary conditions. The domain edge cells are made to be cut cells with volume fraction $\alpha = 10^{-3}$. The simulation is run for 1 period.

\begin{figure}[H]
\centering
\input{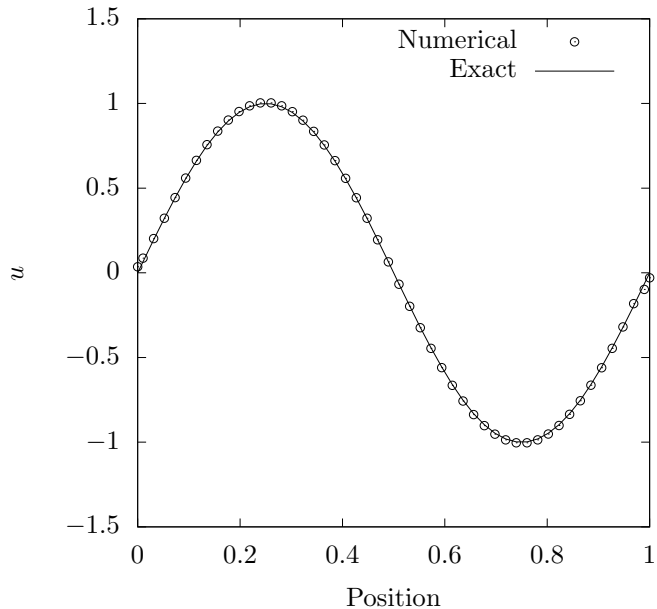}
\caption{Comparison of numerical and exact solutions for the one-dimensional advection problem run with a resolution of 50 cells.}
\label{fig:1DAdv_50_u_vs_x}
\end{figure}

\figref{fig:1DAdv_50_u_vs_x} shows the results for a resolution of 50 cells. The errors for various resolutions are shown in \tableref{table:1DAdvErrorNorms}, where it may be seen that all the norms converge with the expected rates. Note that the $L_\infty$ norm is the same as the maximum cut cell error.

\begin{table}[H]
\caption{Error norms and orders of convergence for the one-dimensional advection problem. The $L_\infty$ norm is the same as the maximum cut cell error.}
\label{table:1DAdvErrorNorms}
\centering
\renewcommand{\arraystretch}{1.1}
\begin{tabularx}{\textwidth}{@{}YYYYYYY@{}}
\hline
Resolution & $L_1$ norm & $L_1$ order & $L_2$ norm & $L_2$ order & $L_\infty$ norm & $L_\infty$ order\\
\hline
$50$ & $6.33 \times 10^{-3}$ & - & $9.72 \times 10^{-3}$ & - & $3.56 \times 10^{-2}$ & -\\
$100$ & $1.55 \times 10^{-3}$ & 2.03 & $3.06 \times 10^{-3}$ & 1.67 & $1.85 \times 10^{-2}$ & 0.95\\
$200$ & $3.94 \times 10^{-4}$ & 1.98 & $1.07 \times 10^{-3}$ & 1.51 & $1.00 \times 10^{-2}$ & 0.88\\
$400$ & $1.00 \times 10^{-4}$ & 1.97 & $3.82 \times 10^{-4}$ & 1.49 & $5.29 \times 10^{-3}$ & 0.92\\
\hline
\end{tabularx}
\end{table}

\subsubsection{Two-dimensional diagonal advection}

\begin{figure}[H]
\centering
\subfloat[Numerical contours after 1 period from the $50 \times 50$ cells simulation.]{
\includegraphics[width=0.6\textwidth]{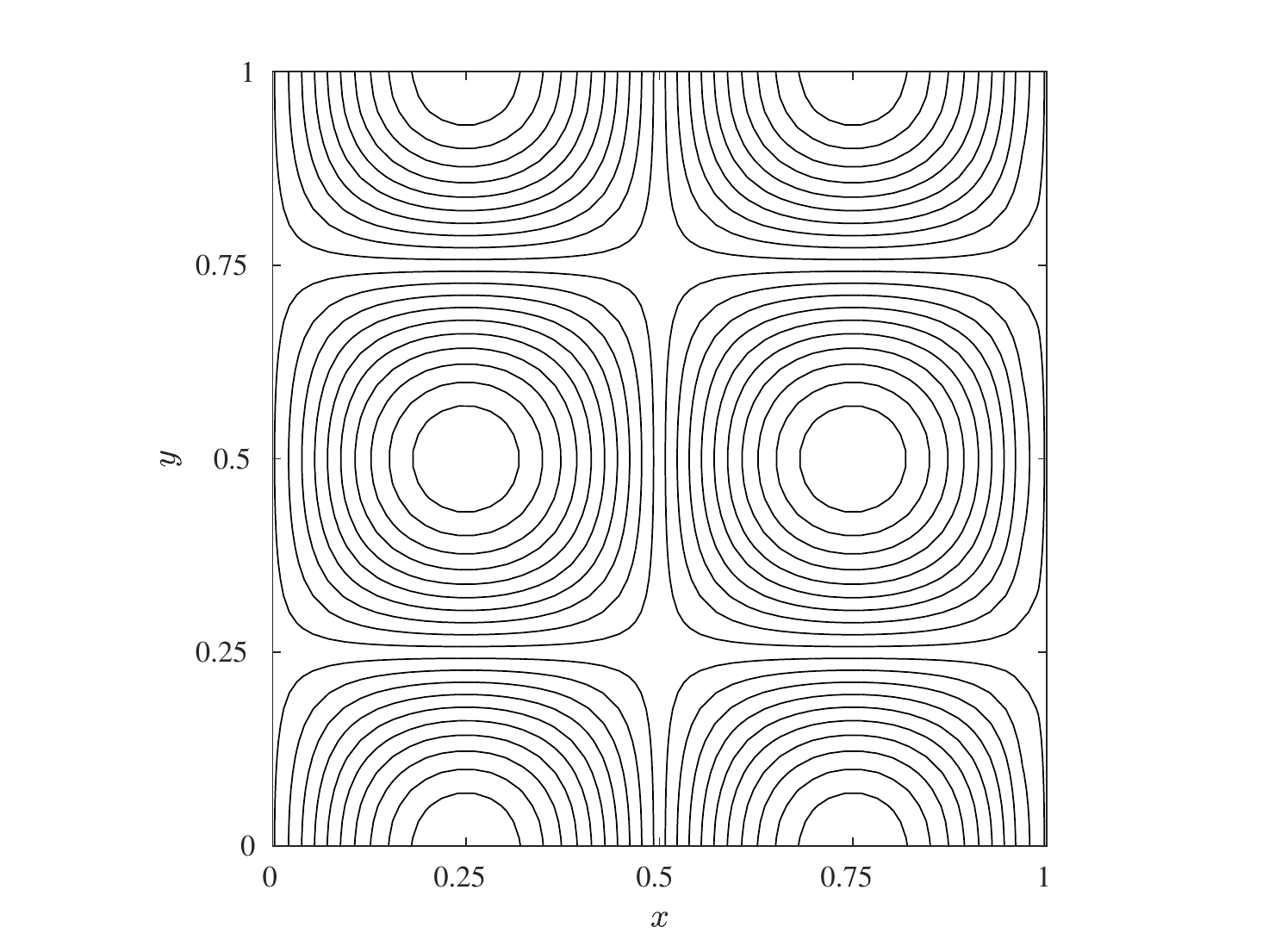}
\label{fig:2DAdv_50_50}}
\:\:\subfloat[Close-up of the top-right part of the mesh. The boundary cut cells are illustrated with an exaggerated volume fraction.]{
\raisebox{0.9cm}{\input{./2DAdvCCMesh.tfig}}
\label{fig:2DAdvCCMesh}}
\caption{Plot of the numerical solution and an illustration of the cut cell mesh for the two-dimensional diagonal advection problem.}
\label{fig:2dAdv_sol_plus_mesh}
\end{figure}

This problem involves the linear advection of the smooth function,
\begin{equation}
u(x,y) = \sin(2\pi x)\cos(2\pi y),
\end{equation}
with a propagation velocity $\mathbf{a} = [1.0 \:\:\: 1.0]^{T}$ in the interval $x \in [0,1], y \in [0,1]$. The boundary conditions are periodic and the domain edge cells are all made to be cut cells with volume fraction $\alpha = 10^{-3}$. As shown in \figref{fig:2DAdvCCMesh}, which is an illustration of the top-right portion of the mesh, the 4 cut cells at the corners of the domain have a volume fraction of $\alpha^2 = 10^{-6}$.

The simulation is run for 1 period. \figref{fig:2DAdv_50_50} shows the numerical contours produced from a $50 \times 50$ cells simulation. \tableref{table:2DAdvErrorNorms} shows the errors, where it may be seen that the norms all converge with the expected rates. Note that the $L_\infty$ norm is the same as the maximum cut cell error.

\begin{table}[H]
\caption{Error norms and orders of convergence for the two-dimensional diagonal advection problem. The $L_\infty$ norm is the same as the maximum cut cell error.}
\label{table:2DAdvErrorNorms}
\centering
\renewcommand{\arraystretch}{1.1}
\begin{tabularx}{\textwidth}{@{}YYYYYYY@{}}
\hline
Resolution & $L_1$ norm & $L_1$ order & $L_2$ norm & $L_2$ order & $L_\infty$ norm & $L_\infty$ order\\
\hline
$50 \times 50$ & $6.44 \times 10^{-3}$ & - & $8.75 \times 10^{-3}$ & - & $3.60 \times 10^{-2}$ & -\\
$100 \times 100$ & $1.56 \times 10^{-3}$ & 2.04 & $2.51 \times 10^{-3}$ & 1.80 & $1.85 \times 10^{-2}$ & 0.96\\
$200 \times 200$ & $3.92 \times 10^{-4}$ & 1.99 & $8.21 \times 10^{-4}$ & 1.61 & $1.00 \times 10^{-2}$ & 0.88\\
$400 \times 400$ & $9.88 \times 10^{-5}$ & 1.99 & $2.81 \times 10^{-4}$ & 1.55 & $5.29 \times 10^{-3}$ & 0.92\\
\hline
\end{tabularx}
\end{table}

\subsubsection{Two-dimensional advection in a sloped channel}

This test involves solving the full Euler system for the parallel advection of a Gaussian density profile in a sloped channel. This is a non-trivial problem for the split LPFS scheme since any errors in the boundary and cut cell flux computations would affect the ability of the method to maintain the parallel flow.

\begin{figure}[H]
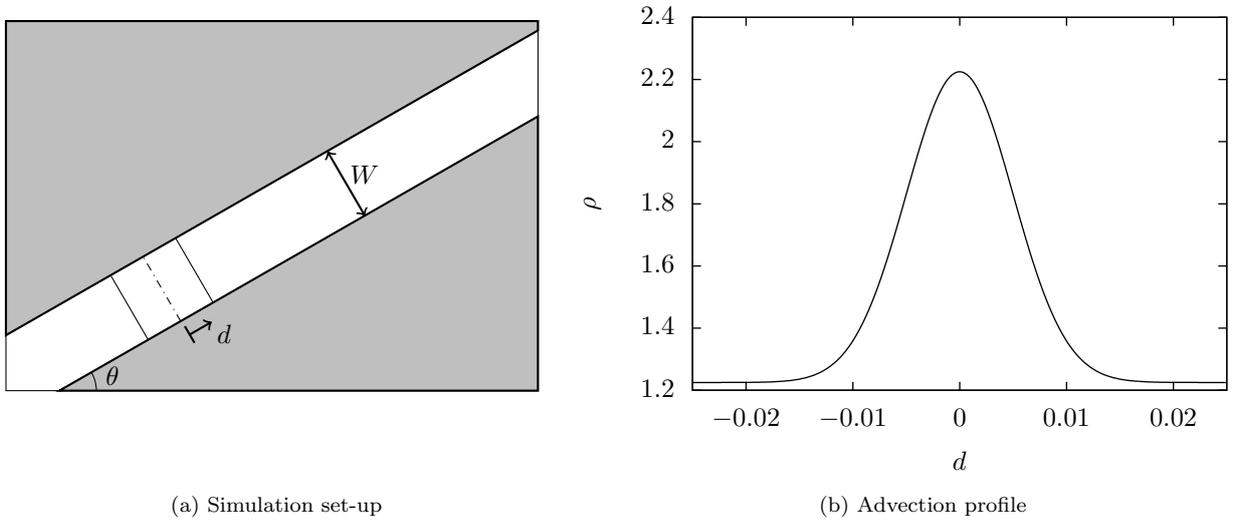

\centering
\subfloat[Simulation set-up]{
\raisebox{1.2cm}{\input{./advectionSetup.tfig}}
\label{fig:advectionSetup}}
\subfloat[Advection profile]{
\input{./advectionProfile.tex}
\label{fig:advectionProfile}}
\caption{Illustration of the simulation set-up and the density profile being advected for the two-dimensional advection in a sloped channel problem.}
\label{fig:advection_setup_plus_profile}
\end{figure}

As illustrated in \figref{fig:advectionSetup}, the channel with a width $W = 0.0141$ m makes an angle $\theta = 30^\circ$ with the $x$ axis. The pressure and velocity are $101325 \: \text{Pa}$ and $30 \:  \text{m/s}$ parallel to the channel wall respectively. The density profile is shown in \figref{fig:advectionProfile} and described by the following equation:
\begin{equation}
\label{eqn:advectionProfile}
\rho(d) = \rho_0 + e^{-\left( \frac{d}{0.5W} \right)^2},
\end{equation}
where $\rho_0 = 1.225 \: \text{kg}/\text{m}^{3}$ and $d$ is distance measured from the centre of the profile in the direction parallel to the channel wall.

\begin{figure}[H]
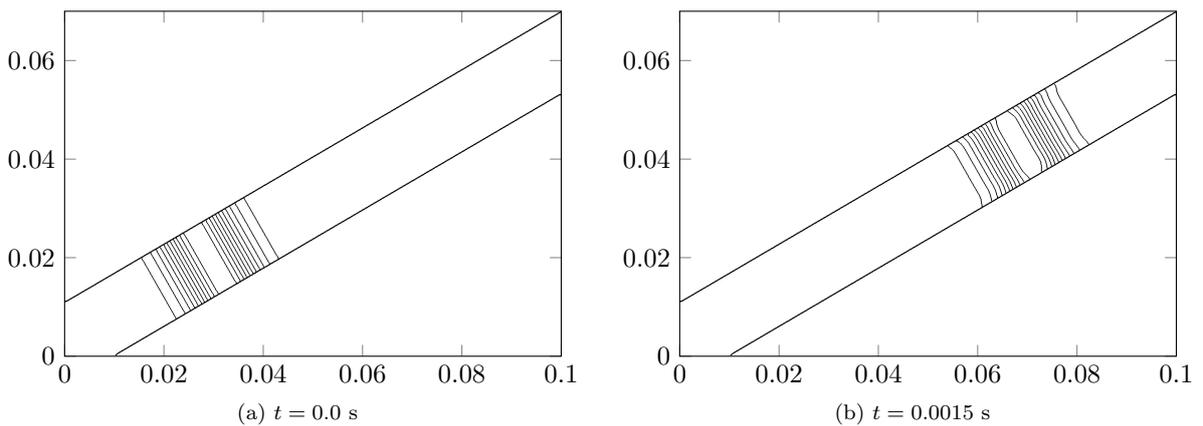

\centering
\subfloat[$t = 0.0$ s]{
\input{./adv2D_contours_t0_0.tfig}
\label{fig:adv2D_contours_t0_0}}
\subfloat[$t = 0.0015$ s]{
\input{./adv2D_contours_t0_0015.tfig}
\label{fig:adv2D_contours_t0_0015}}
\caption{Density contours at the start and end of the simulation for the two-dimensional advection in a sloped channel problem run at a resolution of $200 \times 140$ cells.}
\label{fig:adv2D_contours}
\end{figure}

The domain size is $[0.0,0.1]\:\text{m} \times [0.0,0.07]\:\text{m}$. At $t = 0$ s, we position the centre of the profile at the point [0.035 cos($\theta$), 0.035 sin($\theta$)] m. The simulation is run till $t = 0.0015$ s, by which time, as seen from \figref{fig:adv2D_contours}, the profile has traversed a large portion of the channel and encountered a range of boundary cut cells.

\begin{figure}[H]
\centering
\subfloat[Lower boundary.]{
\input{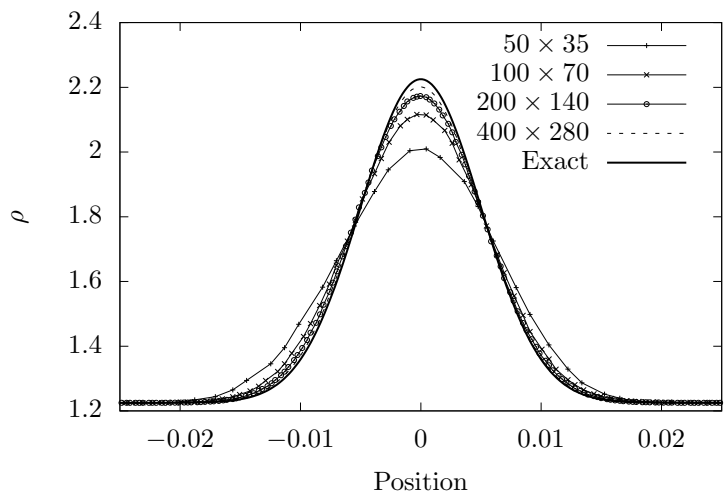}
\label{fig:2DAdv_LowerPlate}}
\\
\subfloat[Upper boundary.]{
\input{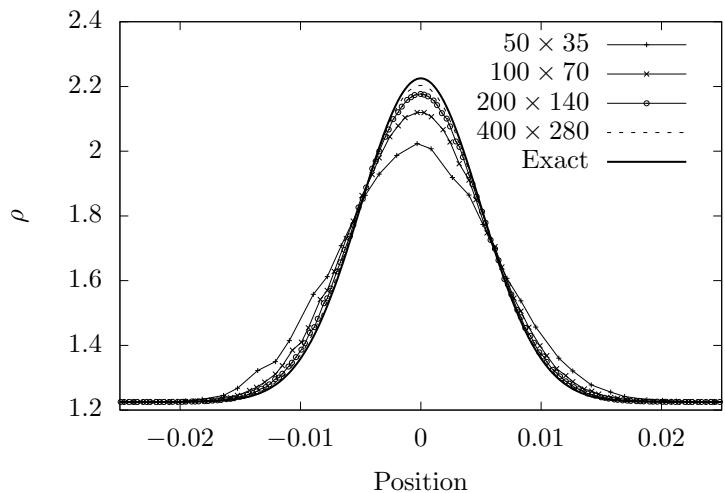}
\label{fig:2DAdv_UpperPlate}}
\caption{Comparison of the numerical solutions at various resolutions with the exact solution along the cut cell boundaries for the two-dimensional advection in a sloped channel problem.}
\label{fig:2dAdvChannelCCSolution}
\end{figure}

\tableref{table:2DAdvChannelErrorNorms} shows the errors computed for simulations at various resolutions. With increasing resolution, it may be observed that the norms converge with the expected rates. Note that the $L_\infty$ norm is the same as the maximum cut cell error and that it converges with first order as expected. \figref{fig:2dAdvChannelCCSolution} shows the final numerical solution along the lower and upper cut cell boundaries. The convergence of the results towards the exact solution with increasing resolution is readily observed.

\begin{table}
\caption{Error norms and experimental orders of convergence for the two-dimensional advection in a sloped channel problem. The $L_\infty$ norm is the same as the maximum cut cell error.}
\label{table:2DAdvChannelErrorNorms}
\centering
\renewcommand{\arraystretch}{1.1}
\begin{tabularx}{\textwidth}{@{}YYYYYYY@{}}
\hline
Resolution & $L_1$ norm & $L_1$ order & $L_2$ norm & $L_2$ order & $L_\infty$ norm & $L_\infty$ order\\
\hline
$50 \times 35$ & $1.97 \times 10^{-2}$ & - & $4.59 \times 10^{-2}$ & - & $2.63 \times 10^{-1}$ & -\\
$100 \times 70$ & $6.19 \times 10^{-3}$ & 1.67 & $1.69 \times 10^{-2}$ & 1.44 & $1.32 \times 10^{-1}$ & 0.99\\
$200 \times 140$ & $1.72 \times 10^{-3}$ & 1.84 & $5.75 \times 10^{-3}$ & 1.56 & $6.09 \times 10^{-2}$ & 1.12\\
$400 \times 280$ & $4.67 \times 10^{-4}$ & 1.89 & $1.98 \times 10^{-3}$ & 1.54 & $2.92 \times 10^{-2}$ & 1.06\\
\hline
\end{tabularx}
\end{table}

\begin{figure}[H]
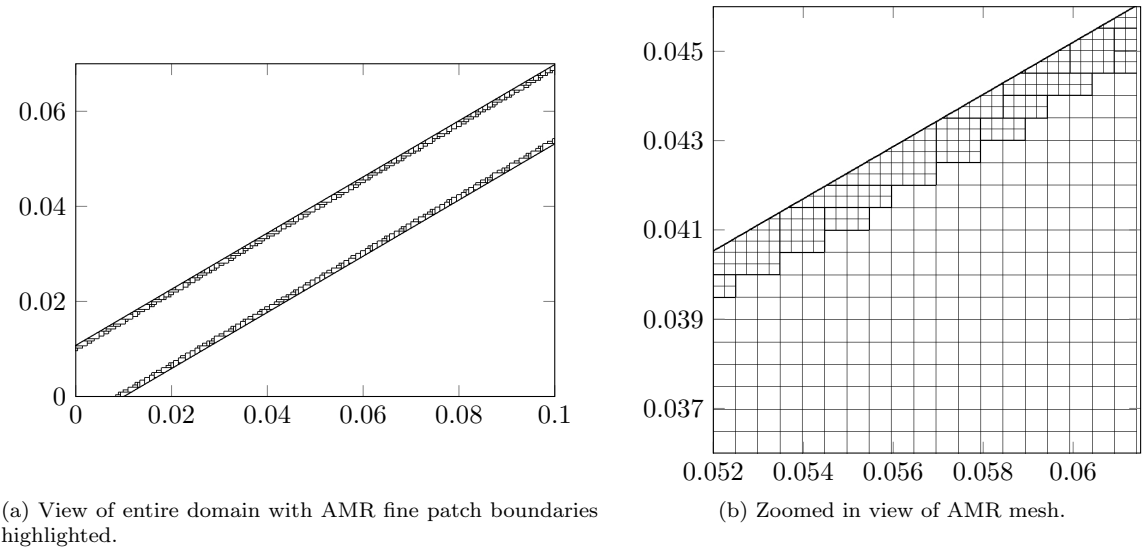

\centering
\subfloat[View of entire domain with AMR fine patch boundaries highlighted.]{
\raisebox{0.75cm}{\input{./adv2D_CCFlagging.tfig}}
\label{fig:adv2D_CCFlagging}}
\quad
\subfloat[Zoomed in view of AMR mesh.]{
\input{./adv2D_AMRZoom.tfig}
\label{fig:adv2D_CCFlagging}}
\caption{Plots showing the use of AMR to refine around cut cells for the two-dimensional advection in a sloped channel problem. The base resolution is $200 \times 140$ cells.}
\label{fig:adv2D_AMRMesh}
\end{figure}

It is also useful to note that a judicious use of AMR can be used to alleviate the effect of reduced order of accuracy at the boundary. \tableref{table:advectionAMRErrorNorms} shows the errors for a series of closely related simulations. The first row shows the error norms for a $100 \times 70$ cells simulation with no AMR. Using these results as a reference, the second row shows the reduced norms that would be theoretically expected from a $200 \times 140$ cells simulation if we had a universally second order method. The errors obtained in practice with the LPFS method are clearly larger, as shown in the third row. The fourth row shows the norms from a $200 \times 140$ cells simulation where one level of AMR of refinement factor 2 is used to refine the cut cells, as shown in \figref{fig:adv2D_AMRMesh}. When compared to the theoretical results of the second row, the error norms for this set-up are in fact lower. Of course, we do not lose sight of the fact that the cut cells from the AMR simulation are at an `effective' resolution of $400 \times 280$ cells. The intention here is merely to illustrate how AMR can be used to alleviate the effect of the method being first order at the boundary.

\begin{table}
\caption{Data illustrating the use of AMR to alleviate the effect of the cut cell method being first order at the boundaries.}
\label{table:advectionAMRErrorNorms}
\centering
\renewcommand{\arraystretch}{1.1}
\begin{tabularx}{\textwidth}{@{}lYYY@{}}
\hline
Resolution & $L_1$ norm & $L_2$ norm & $L_\infty$ norm\\
\hline
$100 \times 70$ (no AMR) & $6.19 \times 10^{-3}$ & $1.69 \times 10^{-2}$ & $1.32 \times 10^{-1}$\\
$200 \times 140$ (second order, expected) & $1.55 \times 10^{-3}$ & $4.23 \times 10^{-3}$ & $3.31 \times 10^{-2}$\\
$200 \times 140$ (no AMR) & $1.72 \times 10^{-3}$ & $5.75 \times 10^{-3}$ & $6.09 \times 10^{-2}$\\
$200 \times 140$ (with cut cells refinement) & $1.31 \times 10^{-3}$ & $3.62 \times 10^{-3}$ & $2.93 \times 10^{-2}$\\
\hline
\end{tabularx}
\end{table}

\subsection{Shock reflection from a wedge}

The $M=1.7$ shock reflection off a $30^\circ$ wedge test problem from Toro \cite{Toro2009} is used to demonstrate the improved performance at the boundary of the LPFS flux compared to the KBN flux. The ambient state ahead of the shock has a density and pressure of 1.225 $\text{kg}/\text{m}^{3}$ and 101325 Pa respectively. The domain size was $[0.0,16.5]\:\text{m} \times [0.0,25.0]\:\text{m}$. A base resolution of $500 \times 330$ cells was used and two levels of AMR refinement of factor 2 each were employed to resolve the shocks and slip line. The boundary conditions were transmissive at the left, right and top boundaries, and reflective at the bottom boundary.

\begin{figure}
\centering
\subfloat[Experimental shadowgraph \cite{Toro2009}.]{
\includegraphics[width=0.45\textwidth]{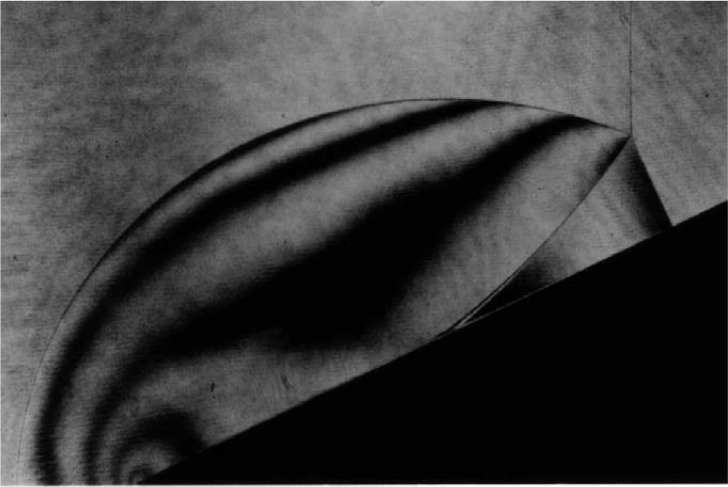}
\label{fig:sMRefl_exp}}
\quad \quad
\subfloat[Numerical density contours.]{
\includegraphics[width=0.456\textwidth]{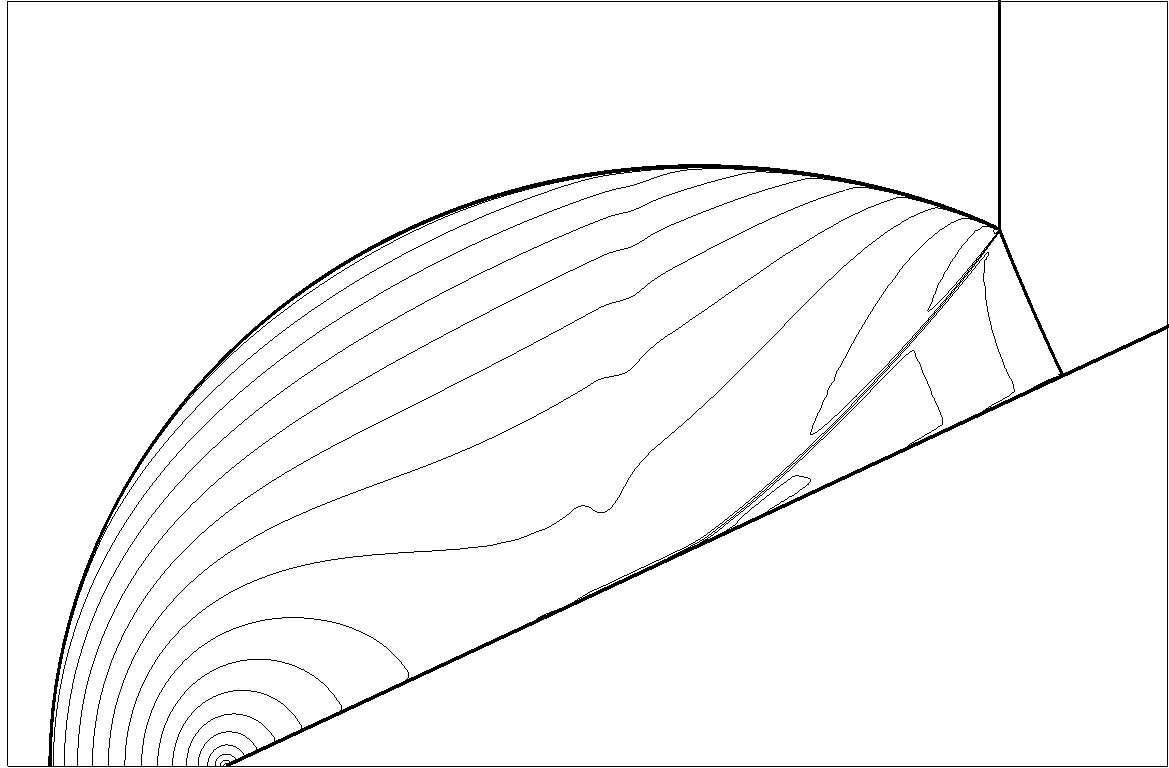}
\label{fig:sMRefl_LPFS_contours}}
\caption{Comparison of experimental and LPFS simulation results for the shock reflection from wedge problem.}
\label{fig:sMRefl_exp_vs_LPFS}
\end{figure}

\figref{fig:sMRefl_exp_vs_LPFS} shows a comparison of the experimental and numerical results. The simulation captures all the expected features. The incident shock, reflected shock, and Mach stem (which is perpendicular to the wedge) meet at the `triple point'. The slip line connecting the triple point to the wedge is also resolved.

\begin{figure}[H]
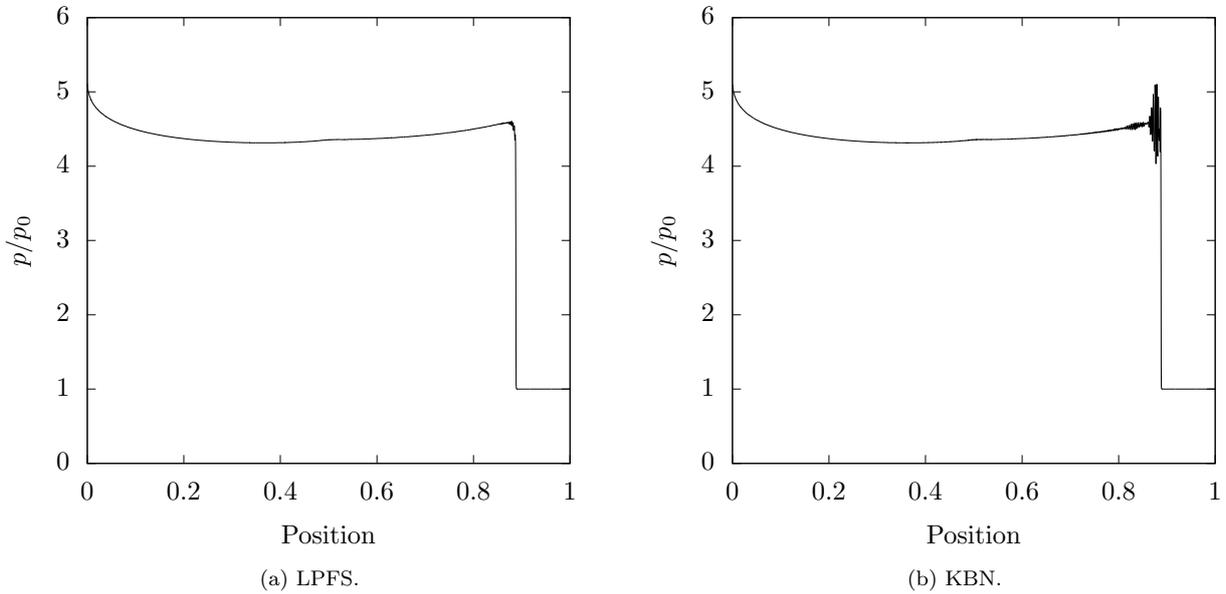

\centering
\subfloat[LPFS.]{
\input{./wedgeSurfPLPFS.tex}
\label{fig:wedgeSurfPLPFS}}
\subfloat[KBN.]{
\input{./wedgeSurfPKBN.tex}
\label{fig:wedgeSurfPKBN}}
\caption{Comparison of the cut cell pressure results for LPFS and KBN for the shock reflection from wedge problem. The pressure is plotted against the normalised distance along the length of the wedge. $p_0 = 101325$ Pa.}
\label{fig:wedgeSurfP}
\end{figure}

\figref{fig:wedgeSurfP} shows a comparison of the surface pressure distribution computed using the LPFS and KBN fluxes. The KBN solution behind the Mach stem is highly oscillatory, and the issue is greatly alleviated in the LPFS solution. Reducing the Courant number to say, 0.5, would improve the KBN solution, however the intention of this test is to demonstrate the superior accuracy and robustness of LPFS at higher Courant numbers.

\subsection{Subsonic flow over a NACA 0012 aerofoil}

The computation of $M=0.6$ subsonic flow over a symmetric NACA 0012 aerofoil with 0 angle of attack is used to demonstrate the increased accuracy of LPFS compared to KBN near stagnation points.

The aerofoil with a chord length $c = 0.127$ m is placed at approximately the centre of a domain having a length and height of $30c$. A coarse base resolution of $200 \times 200$ cells is used to help accelerate convergence to steady state by diffusing waves bouncing off the domain edge boundaries. The finest resolution simulation employed four AMR levels with refinement factors of 4, 4, 4 and 2. The free stream density $\rho_\infty$ and pressure $p_\infty$ are 1.225 $\text{kg}/\text{m}^{3}$ and 101325 Pa respectively. Subsonic inflow boundary conditions are used at the left boundary with outflow conditions specified at the right, top and bottom boundaries respectively.

\begin{figure}[H]
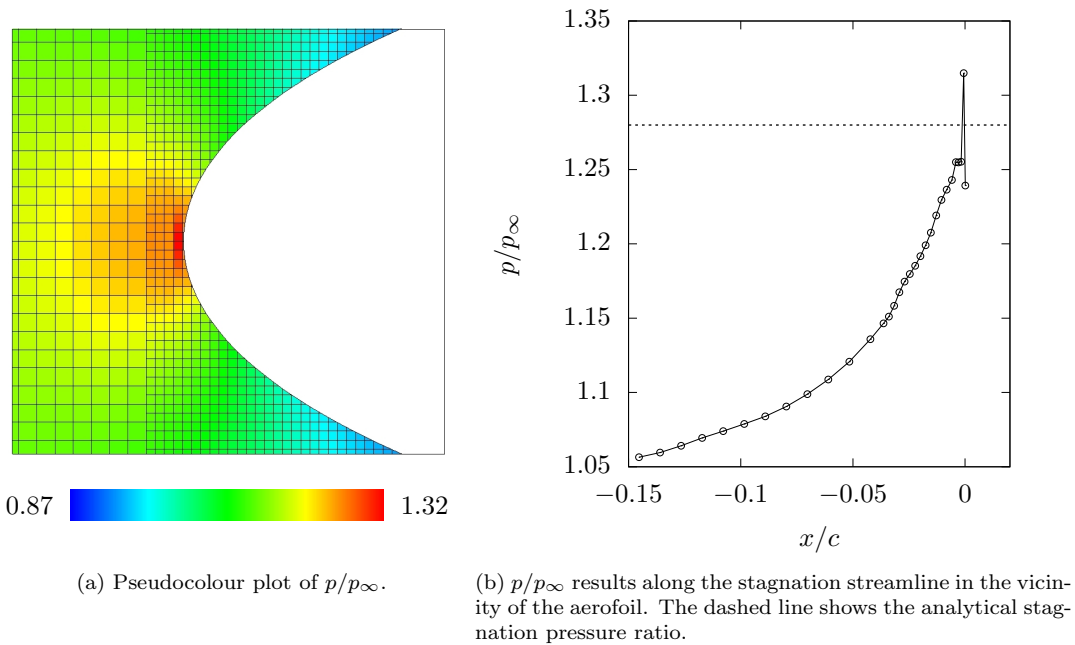

\centering
\subfloat[Pseudocolour plot of $p/p_\infty$.]{
\raisebox{0.5cm}{\input{./NACA_KBN_pContours.tfig}}
\label{fig:NACA_KBN_pContours}}
\subfloat[$p/p_\infty$ results along the stagnation streamline in the vicinity of the aerofoil. The dashed line shows the analytical stagnation pressure ratio.]{
\input{./NACA_KBN_stagP.tex}
\label{fig:NACA_KBN_stagP}}
\caption{Pressure results with the KBN flux in the vicinity of the stagnation region for the subsonic NACA 0012 problem.}
\label{fig:NACA_KBN_stagP_Results}
\end{figure}

\figref{fig:NACA_KBN_pContours} is a pseudocolour plot of the pressure ratio $p/p_\infty$ near the stagnation point for the KBN simulation. \figref{fig:NACA_KBN_stagP} shows the pressure ratio along the stagnation streamline in the vicinity of the nose of the aerofoil. The dashed line corresponds to the theoretical stagnation pressure ratio calculated from the isentropic flow relations \cite{Anderson2010}. Clearly, the pressure computed with the KBN flux is too low in the stagnation cut cell and too high in the regular neighbouring cell.

\begin{figure}[H]
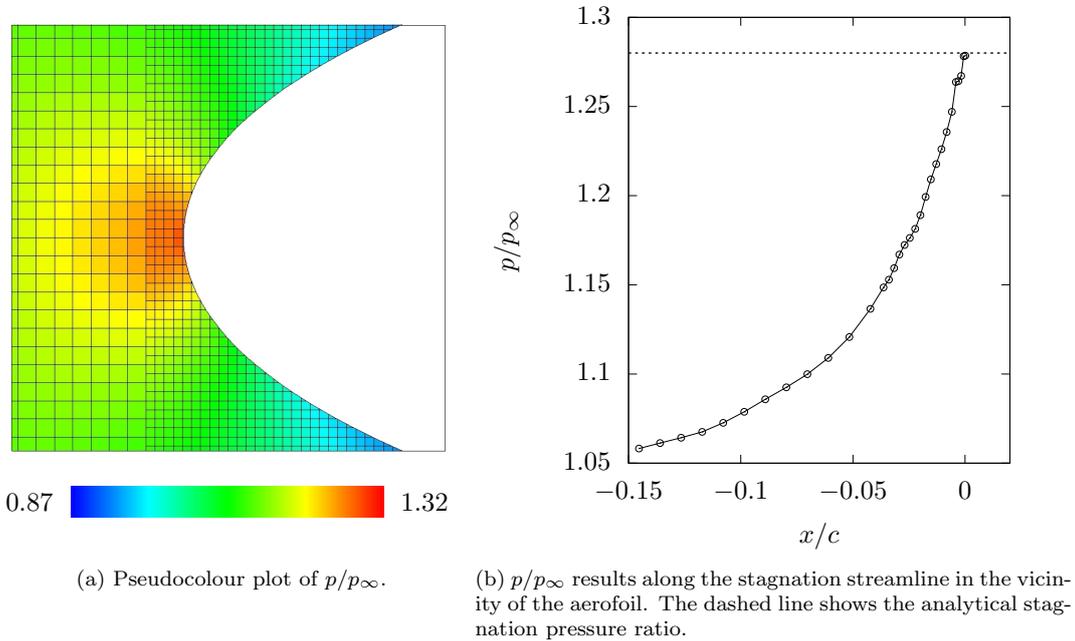

\centering
\subfloat[Pseudocolour plot of $p/p_\infty$.]{
\raisebox{0.5cm}{\input{./NACA_LPFS_pContours.tfig}}
\label{fig:NACA_LPFS_pContours}}
\subfloat[$p/p_\infty$ results along the stagnation streamline in the vicinity of the aerofoil. The dashed line shows the analytical stagnation pressure ratio.]{
\input{./NACA_LPFS_stagP.tex}
\label{fig:NACA_LPFS_stagP}}
\caption{Pressure results with the LPFS flux in the vicinity of the stagnation region for the subsonic NACA 0012 problem.}
\label{fig:NACA_LPFS_stagP_Results}
\end{figure}

\figref{fig:NACA_LPFS_stagP_Results} shows the corresponding results computed with the LPFS flux. The solution looks visibly improved in the pseudocolour plot of \figref{fig:NACA_LPFS_pContours}, where the limits for the colour bar have been set to be the same as those in \figref{fig:NACA_KBN_pContours}. \figref{fig:NACA_LPFS_stagP} confirms that the LPFS stagnation solution shows good agreement with the analytical solution. These results confirm that the introduction of local wave speeds in the LPFS flux stabilisation produces the intended effect.

\begin{figure}[H]
\centering
\subfloat[Experimental \cite{Harris} vs numerical surface pressure distributions.]{
\input{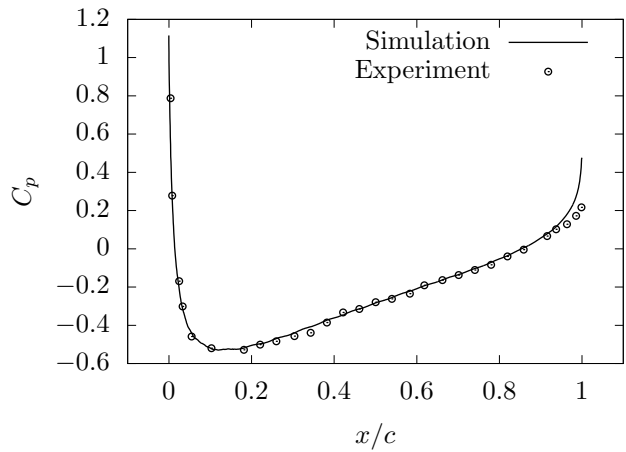}
\label{fig:NACA_LPFS_Cp}}
\subfloat[Convergence plot of $C_D$ computation.]{
\input{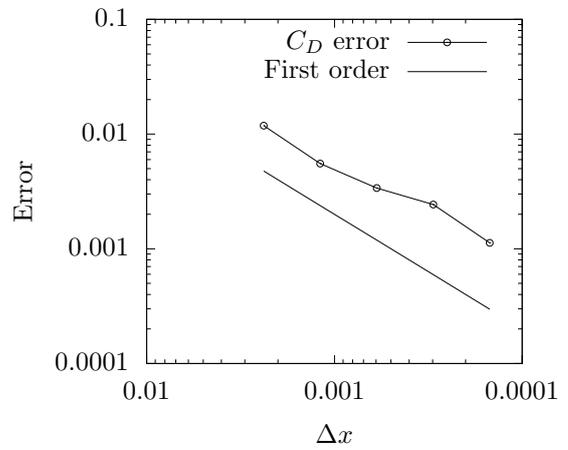}
\label{fig:NACA_Drag_Convergence}}
\caption{Surface pressure and drag results for the subsonic NACA 0012 problem.}
\label{fig:NACA_Surf_Plus_Drag}
\end{figure}

\figref{fig:NACA_LPFS_Cp} shows a comparison of the numerical and experimental (see Harris \cite{Harris}) pressure distributions over the aerofoil. Note that $C_p$ is the non-dimensional pressure coefficient:
\begin{equation}
\label{eqn:Cp}
C_p = \frac{p-p_\infty}{\frac{1}{2} \rho_\infty u_\infty^2},
\end{equation}
where $u_\infty$ is the free stream velocity. The computed solution shows good agreement with the experimental measurements over the whole of the aerofoil.

\figref{fig:NACA_Drag_Convergence} shows how the error in the computed drag coefficient, $C_D$, decreases with increasing resolution. Note that
\begin{equation}
\label{eqn:CD}
C_D = \frac{F_D}{\frac{1}{2} \rho_\infty u_\infty^2 c},
\end{equation}
where $F_D$ is the pressure drag force. The theoretically expected drag force for this non-separating inviscid flow is 0, and any computed positive $C_D$ is a measure of the discretisation error. \figref{fig:NACA_Drag_Convergence} shows a first order convergence for $C_D$ which is in line with expectations since the cut cell method is first order accurate at the boundary.

\subsection{Shock reflection over a double wedge}

\label{sect:Shock_reflection_over_a_double_wedge}

\begin{figure}[H]
\centering
\input{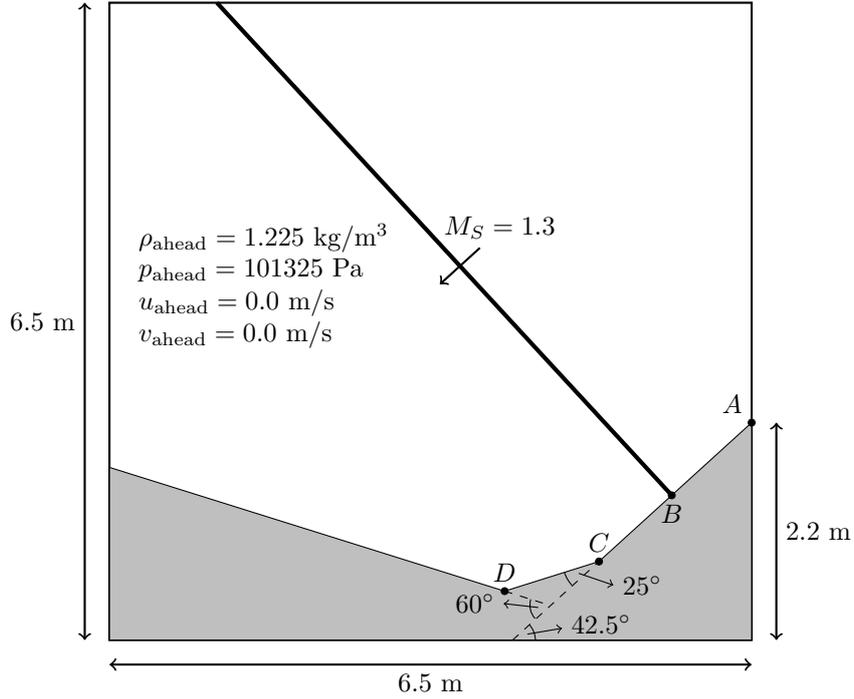}
\caption{Rotated simulation set-up for the problem of shock reflection over a double wedge. Note that $AB=BC=CD=1$ m.}
\label{fig:rot_doubleWedgeSetup}
\end{figure}

The problem of a $M=1.3$ normal shock reflecting over a double wedge is used to demonstrate the performance of the method for a 2D problem involving concavities. The simulation set-up is shown in \figref{fig:rot_doubleWedgeSetup}, where it may be seen that the wedge and shock have been rotated in order to create a fully doubly-shielded concavity (see \sectionref{sect:Post_sweep_correction_at_concavities}) with respect to the grid at point $D$. Note that points $A$ and $B$ are located upstream of the wedge leading edge, which is at point $C$. To resolve the detailed solution structure, a fine base resolution of $1200 \times 1200$ cells is employed with two levels of AMR refinement of factor 2 each. Boundary conditions on the left, right and top boundaries are all transmissive.

\begin{figure}[H]
\centering
\subfloat[Mach reflection over the second wedge of the Mach stem from the first wedge reflection.]{
\includegraphics[width=0.35\textwidth]{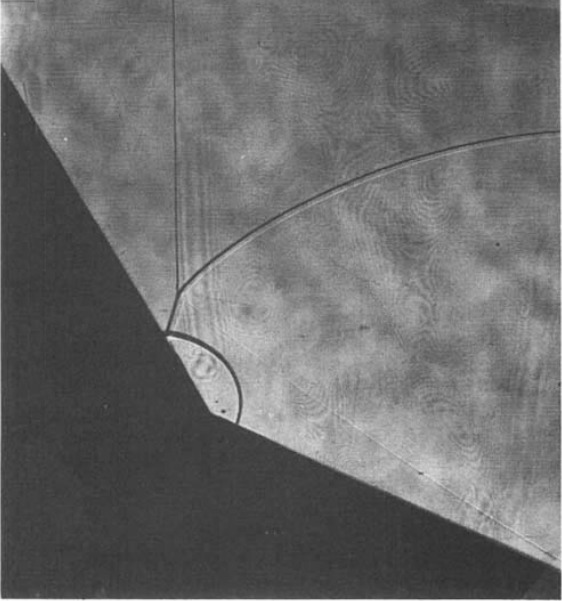}
\label{fig:expShadow1}}
\quad \quad
\subfloat[Final flow structure.]{
\includegraphics[width=0.35\textwidth]{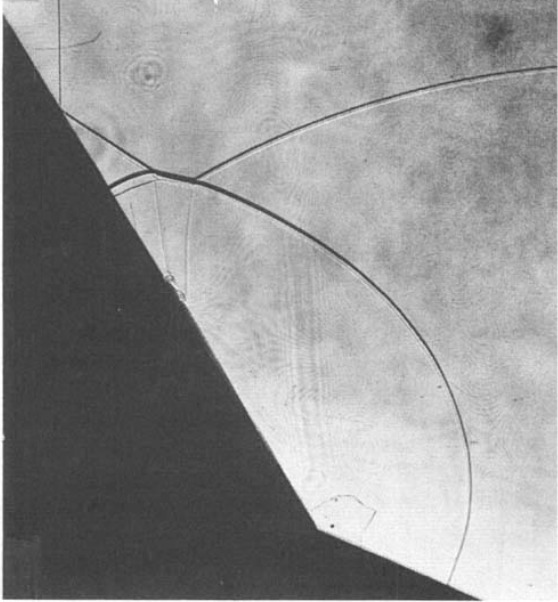}
\label{fig:expShadow2}}
\caption{Experimental shadowgraphs \cite{Ben-Dor1987} for the shock reflection over a double wedge problem.}
\label{fig:dWedgeExp}
\end{figure}

\begin{figure}[H]
\centering
\subfloat[Solution at $t=0.0044$ s.]{
\includegraphics[width=0.35\textwidth]{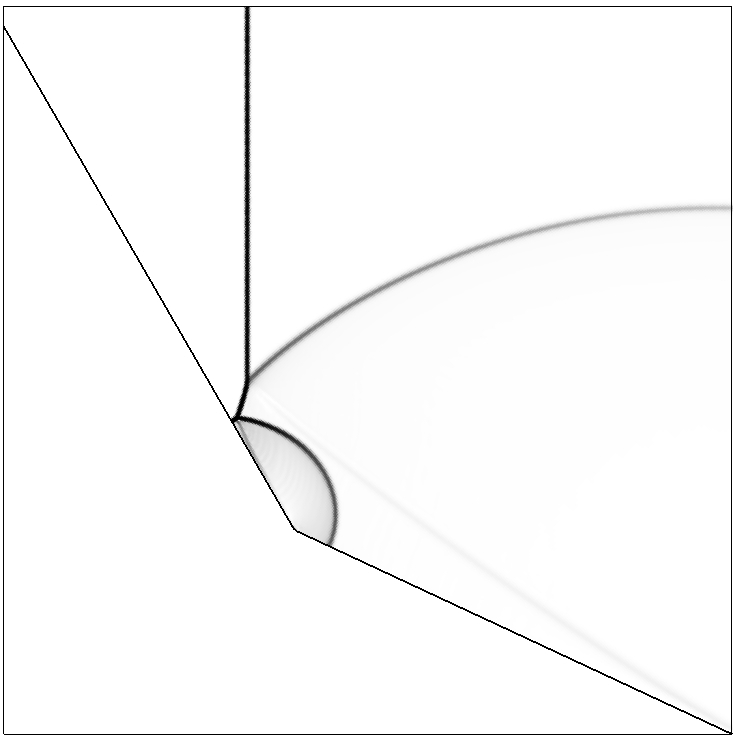}
\label{fig:dWedgeSchlt0_0044}}
\quad\quad
\subfloat[Solution at $t=0.005$ s.]{
\includegraphics[width=0.35\textwidth]{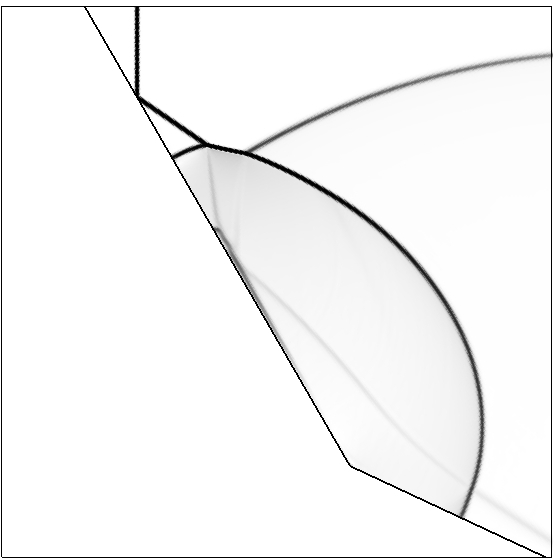}
\label{fig:dWedgeSchlt0_005}}
\caption{Numerical schlierens for the shock reflection over a double wedge problem.}
\label{fig:dWedgeSchl}
\end{figure}

\figref{fig:dWedgeExp} shows experimental shadowgraphs for this test as measured by Ben-Dor, Dewey and Takayama \cite{Ben-Dor1987}, who provide detailed explanations of the wave interactions that lead to the observed solution structures. Numerical schlierens of the computed density field at $t = 0.0044$ s and $0.005$ s are shown in \figref{fig:dWedgeSchlt0_0044} and \figref{fig:dWedgeSchlt0_005} respectively. For ease of comparison, note that we rotate our results to align them with the experimental frame.

The numerical results show good qualitative agreement with experiment at both times. From \figref{fig:dWedgeSchlt0_0044}, we see that the simulation captures the Mach reflection over the second wedge of the Mach stem from the first wedge reflection. The slip line from the first Mach reflection is also clearly resolved. Noteworthy in \figref{fig:dWedgeSchlt0_005} is the resolution of the two triple points and their slip lines. Another feature captured by the AMR is the interaction of the slip lines with the contact line from the first Mach reflection. This feature is not discussed by Ben-Dor et al. but may be seen to be just visible in the experimental shadowgraph.

\subsection{Shock diffraction over a cone}

A full 3D simulation of the diffraction of a $M=3.55$ normal shock over a cone is used to validate the three-dimensional implementation of LPFS. Incidentally, this test has also been used by Yang, Causon and Ingram \cite{Yang2000} to validate a cell-merging based cut cell method.

The right cone of semi-apex angle $35.1^\circ$ and length 2.0 m has its apex situated at the point $(0,0,0)\:\text{m}$ in a $[-2.5,0.6]\:\text{m} \times [-2.0,2.0]\:\text{m} \times [-2.0,2.0]\:\text{m}$ domain. A base resolution of $62 \times 80 \times 80$ cells is employed with two AMR levels of refinement factors 4 and 2 respectively. The ambient state ahead of the shock has a density and pressure of 1.225 $\text{kg}/\text{m}^{3}$, 101325 Pa respectively.

\begin{figure}[H]
\centering
\subfloat[Experimental schlieren \cite{Bryson1961}. I.S. is the `incident shock', R.S. is the `reflected shock', M.S. is the `Mach stem', C.D. is the `contact discontinuity', and T.P. is the `triple point'.]{
\includegraphics[width=0.4\textwidth,trim=0cm -4.25cm 0cm 0cm]{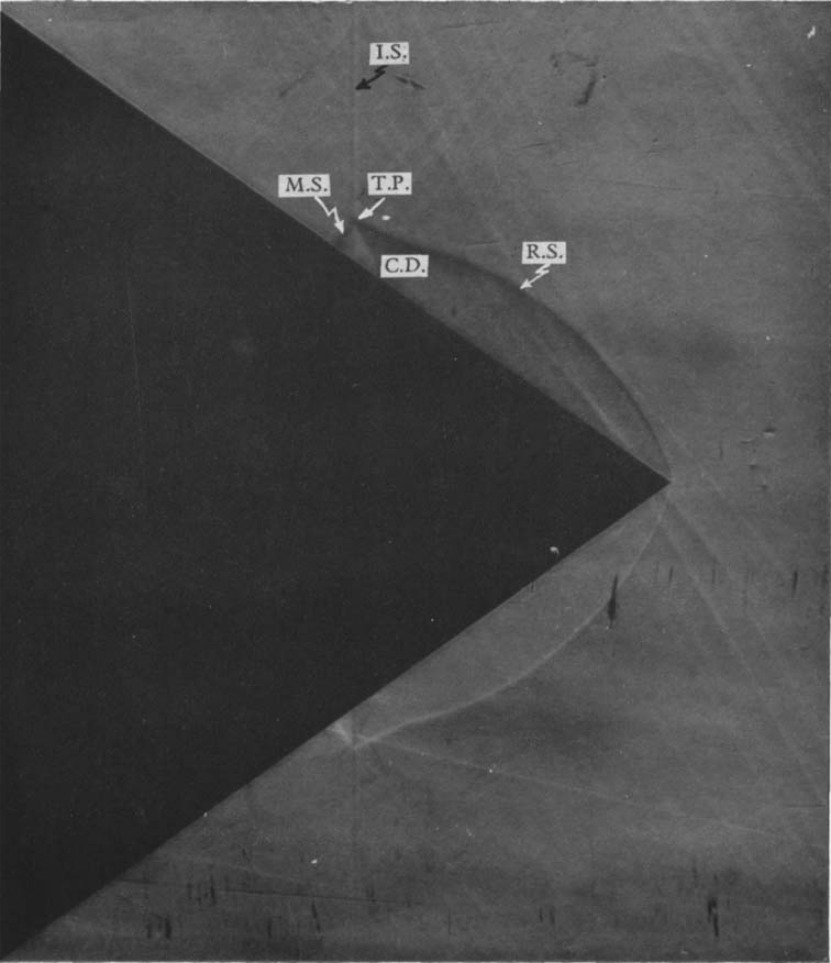}
\label{fig:coneShockDiffExp}}
\quad \quad \quad
\subfloat[Numerical density contours with AMR patch boundaries on the finest level highlighted.]{
\input{./coneShockDiffContours.tfig}
\label{fig:coneShockDiffContours}}

\subfloat[Numerical density contours on the $x$-$y$ and $x$-$z$ symmetry planes.]{
\input{./coneShockDiffContoursz0y0Planes.tfig}
\label{fig:coneShockDiffContoursz0y0Planes}}
\label{fig:coneShockDiffResults}
\caption{Comparison of experimental and numerical results for the shock diffraction over a cone problem.}
\end{figure}

The experimental schlieren measured by Bryson and Gross \cite{Bryson1961} is shown in \figref{fig:coneShockDiffExp}. \figref{fig:coneShockDiffContours} shows the computed density contours on the two symmetry planes ($x$-$y$ and $x$-$z$) of the geometry. All the waves as well as the cut cell boundary are flagged for refinement by the AMR algorithm. The density contours on the symmetry planes are shown more clearly in \figref{fig:coneShockDiffContoursz0y0Planes} which shows that the simulation successfully captures the complex three-dimensional Mach reflection pattern. The numerically resolved incident shock, reflected shock, and Mach stem meeting at the triple point are clearly visible, as is the contact wave.

\subsection{Space re-entry vehicle simulation}

\label{sect:Space_reentry_vehicle_simulation}

\begin{figure}[H]
\centering
\subfloat[Space re-entry vehicle geometry rear view.]{
\includegraphics[width=0.35\textwidth]{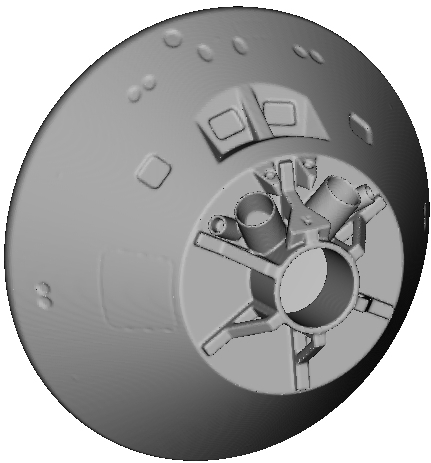}
\label{fig:orionGeom}}
\\
\subfloat[Pseudocolour plot of the natural log of pressure.]{
\includegraphics[width=0.65\textwidth]{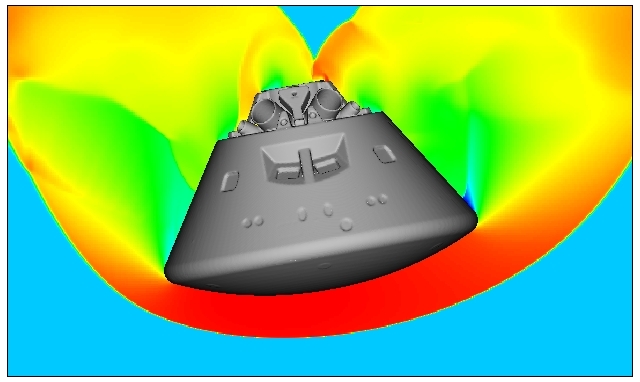}
\label{fig:orionPressure}}
\caption{Geometry rear view and simulation results for the space re-entry vehicle simulation problem.}
\label{fig:orionGeomAndPressure}
\end{figure}

The computation of a $M=20$ flow over a NASA `Orion' space re-entry vehicle is used to demonstrate the performance of the 3D LPFS implementation when computing a high Mach number flow over a realistic complex geometry. A `watertight' stl file of the geometry was obtained from the NASA 3D Resources website \cite{Leon2015}. \figref{fig:orionGeom} shows a view of the geometry from the rear.

In a Mach 20 flow, air molecules undergo dissociation and it is no longer appropriate to use the ideal gas equation of state \eqnref{eqn:idealEOS}. We stress, therefore, that the aim of this section is only to demonstrate the potential of the methodology and not to compute a physically accurate solution. The simulation is set-up with a supersonic inflow boundary condition at the inlet and transmissive boundary conditions at all other boundaries. The geometry is rotated to make an angle of $10^\circ$ with the free stream direction. Since this problem is sensitive to the `carbuncle phenomenon' \cite{Quirk1994a}, we use the HLL Riemann solver to compute fluxes from the reconstructed states. We let the simulation run until the bow shock forms ahead of the spacecraft and till the flow impacts all parts of the geometry. One level of AMR is employed to resolve the solid-fluid interface and the bow shock.

\figref{fig:orionPressure} shows the natural log of the computed pressure along a plane passing through the model centre line. The potential of the methodology to compute the flow around a complex 3D geometry is apparent from the results. Since we are using an ideal gas equation of state, however, we make no attempt to further analyse the complicated flow field.

\section{Conclusions}

\label{sect:Conclusions}

In this paper, we presented a `Local Proportional Flux Stabilisation' (LPFS) approach for computing cut cell fluxes when solving hyperbolic conservation laws, and described its implementation in the dimensionally split framework of Klein et al. \cite{Klein2009}. The approach makes use of local geometric and wave speed information to define a novel stabilised cut cell flux.

The convergence and stability of the method was proved for the one-dimensional linear advection equation, and confirmed numerically for multi-dimensional test problems for the linear advection and Euler equations.

Compared to the `KBN' cut cell flux described by Klein et al., the LPFS flux is designed to give improved accuracy at stagnation points, and this was demonstrated via the computation of a subsonic flow over a NACA 0012 aerofoil. Furthermore, as confirmed from the results of a shock reflection from wedge problem, the LPFS flux was found to alleviate the problem of oscillatory boundary solutions produced by the KBN flux at higher Courant numbers. The performance of the three-dimensional implementation of the method when computing a high Mach number flow over a realistic complex geometry was demonstrated by the computation of a Mach 20 flow over a space re-entry vehicle.

For the future, it is clear that extending the flux stabilisation to maintain second order accuracy at the boundary will yield the greatest improvement in results. The development of a flux to use at concavities which avoids the need to use the current post-sweep conservative correction would also be a useful contribution.

\section*{Acknowledgements}

Nandan Gokhale thanks the Cambridge Commonwealth, European \& International Trust for financially supporting his research. Rupert Klein acknowledges support by Deutsche Forschungsgemeinschaft through Grants SFB 1029 ``TurbIn'', Project C01 and SFB 1114 ``Scaling Cascades in Complex Systems'', Project C01. All 3D simulations for this paper were performed on the \textit{Darwin} High Performance Computing cluster located at the University of Cambridge.

The authors would like to thank Oliver Strickson, Lukas Wutschitz and Murray Cutforth for useful discussions, Alo Roosing for his implementation of the 3D signed distance function calculator, and Philip Blakely for providing support with the parallel AMR code.

\section*{References}

\bibliography{mybibfile}

\end{document}